\documentclass[pra,amsmath,amssymb,amsfonts,nofootinbib]{revtex4}

\usepackage{bm,graphicx,mathrsfs}

\usepackage{amsmath,bbm}
\usepackage{amsfonts,amssymb}
\usepackage{graphicx}
\usepackage{epsfig}
\usepackage{times}
\usepackage{verbatim}

\newcommand{\be}{\begin{equation}}
\newcommand{\ee}{\end{equation}}
\newcommand{\bq}{\begin{eqnarray}}
\newcommand{\eq}{\end{eqnarray}}
\newcommand{\ba}{\begin{align}}
\newcommand{\ea}{\end{align}}

\newcommand{\1}{\mathbbm{1}}
\newcommand{\id}{{\rm id}}

\newcommand{\ket}[1]{\left | \, #1 \right\rangle}
\newcommand{\bra}[1]{\left \langle #1 \, \right |}

\newcommand{\proj}[1]{\ket{#1}\bra{#1}}

\newcommand{\avr}[1]{\left \langle#1 \right \rangle}

\newcommand{\tr}[1]{{\rm tr}\left[{#1}\right]}

\newcommand{\raw}{\rightarrow}

\newcommand{\bR}{\mathbbm{R}}

\newcommand{\bC}{\mathbbm{C}}
\newcommand{\bL}{\mathbbm{L}}

\newcommand{\cH}{\mathcal{H}}
\newcommand{\cE}{\mathcal{E}}

\newcommand{\cL}{\mathcal{L}}

\newcommand{\cA}{\mathcal{A}}
\newcommand{\cS}{\mathcal{S}}

\newcommand{\cM}{\mathcal M}

\newcommand{\cW}{\mathcal W}
\newcommand{\cO}{\mathcal O}

\newcommand{\half}{\frac{1}{2}}

\newcommand{\Ent}{{\rm Ent}}
\newcommand{\Var}{{\rm Var}}

\newtheorem{theorem}{Theorem}
\newtheorem{lemma}[theorem]{Lemma}
\newtheorem{corollary}[theorem]{Corollary}
\newtheorem{proposition}[theorem]{Proposition}
\newtheorem{definition}[theorem]{Definition}

\newtheorem{conjecture}[theorem]{Conjecture}
\def\Proof{\noindent\textsc{Proof:}}
\def\proof{\Proof}
\def\qed{\leavevmode\unskip\penalty9999 \hbox{}\nobreak\hfill
     \quad\hbox{\leavevmode  \hbox to.77778em{%
               \hfil\vrule   \vbox to.675em%
               {\hrule width.6em\vfil\hrule}\vrule\hfil}}
     \par\vskip3pt}
    {\hspace*{\fill}$\Box$\vspace{1.5ex}\par}

\newcommand{\Sp}{\,\,\,\,\,\,}
\newcommand{\no}{\nonumber\\}

\begin{document}

\title{\sc \large Quantum logarithmic Sobolev inequalities and rapid mixing}

\author{Michael J. Kastoryano$^1$ and Kristan Temme$^2$ \vspace{0.1cm} \\}

\address{$^1$ Dahlem Center for Complex Quantum Systems, Freie Universit\"at Berlin, 14195 Berlin, Germany\\
               $^2$ Center for Theoretical Physics, Massachusetts Institute of Technology, Cambridge, MA 02139, USA}

\date{\today}

\begin{abstract}
A family of logarithmic Sobolev inequalities on finite dimensional quantum state spaces is introduced. The framework of non-commutative $\bL_p$-spaces is reviewed and the relationship between quantum logarithmic Sobolev inequalities and the hypercontractivity of quantum semigroups is discussed. This relationship is central for the derivation of lower bounds for the logarithmic Sobolev (LS) constants. Essential results for the family of inequalities are proved, and we show an upper bound to the generalized LS constant in terms of the spectral gap of the generator of the semigroup. These inequalities provide a framework for the derivation of improved bounds on the convergence time of quantum dynamical semigroups, when the LS constant and the spectral gap are of the same order. Convergence bounds on finite dimensional state spaces are particularly relevant for the field of quantum information theory. We provide a number of examples, where improved bounds on the mixing time of several semigroups are obtained; including the depolarizing semigroup and quantum expanders.
\end{abstract}

\maketitle

\section{Introduction}\label{sec:Intro}
Logarithmic Sobolev inequalities were originally introduced by Gross in 1975 \cite{Gross1,Gross2}, who related them to 
the hypercontractivity of semigroups. Initially, the main focus was on the investigation of logarithmic Sobolev (LS) inequalities, or in short Log-Sobolev inequalities, on infinite dimensional state spaces, until Diaconis and Saloff-Coste used these inequalities to bound the $\bL_1$ - mixing time of finite dimensional classical Markov processes \cite{Diaconis}. A tantalizing example of where these inequalities have given rise to some of the tightest known mixing time bounds for continuous time Markov processes is in the analysis of Ising-type spin systems under Glauber dynamics (see \cite{Martinelli,Guillonet} for more detail).\\
In this paper we generalize the mixing time bounds based on logarithmic Sobolev inequalities to finite dimensional quantum (i.e. non-commutative) state spaces. We consider completely-positive trace-preserving semigroups in continuous time, described by time-independent generators, which can always be cast in Lindblad normal form \cite{Lindblad}. Quantum generalizations of Log-Sobolev inequalities on infinite dimensional $C^*$ algebras have already been considered in \cite{Lieb} for a specific unital fermionic semigroup and were later generalized \cite{Ziggy1} to arbitrary reversible semigroups to investigate hypercontractivity in non-commutative $\bL_p$ spaces \cite{LpSpaces1,LpSpaces2}. Here, we will work exclusively on finite dimensional state spaces and derive bounds on the trace-norm or $\bL_1$ -norm distance between the steady state and the non-equilibrium state of the quantum Markov processes.\\
A central motivation for studying the mixing time behavior of quantum mechanical semigroup stems from the field of quantum information theory, where several questions relate to problems of estimating the time scales of dissipative processes. Indeed, a prime example is the study of decoherence \cite{Schlossauer} of extended quantum systems, where in particular one would like to construct realistic physical systems, that can retain quantum information for long times \cite{terhal}. The central question in this and other studies is how the time to reach equilibrium scales in the system size. Other applications can be found in the investigation of the run times of quantum algorithms based on quantum Markov processes \cite{Frank, Metropolis}. Furthermore, the derivation of rigorous bounds on the thermalization time of quantum mechanical systems poses a central problem in the endeavor
of understanding statistical mechanics from the microscopic quantum theory \cite{thermal}.\\   
Before we proceed with the formal exposition of the subject, let us first consider a simple example, which already illustrates the possible benefit of bounding the mixing time of quantum Markov processes in terms of the Log-Sobolev constant rather than with other figures of merit, such as the spectral gap.\\  

\paragraph{\bf Motivation : }
We assume some familiarity with the standard notation of quantum dynamical semigroups in this section \cite{Breuer}.  
The full formal framework will be introduced in the next section.

The mixing time of a quantum Markov process is the time it takes for the process to become close to the stationary state, starting from any initial state. The distance between two states is usually measured in terms of the trace norm, $\| A \|_{tr} = \tr{|A|}$, since it possesses the appropriate operational interpretation as a distinguishability measure \cite{Fuchs}. Let  $\sigma$ be the stationary state of a semigroup generated by the quantum dynamical master equation  $\partial_t \rho_t = \cL^*(\rho_t)$, where $\cL$ is the Liouvillian in the Heisenberg picture. Then, the mixing time is defined as 
\be
 \tau_{mix}(\epsilon) = \min\left\{ t \left| \|\rho_t - \sigma\|_{tr} \leq \epsilon \; \mbox{for all input states} \;  \rho_0 \right. \right\}
\ee 
Bounds to the trace norm distance for quantum processes have been derived \cite{chi2} in terms of the spectral properties 
of the generators. If the semigroup has a unique full rank stationary state, a general upper bound on the trace distance can be obtained:  
\be\label{chi2Mix}
  \left\| \rho_t - \sigma \right \|_{tr} \leq \sqrt{1/\sigma_{\min}}e^{-\lambda t},
\ee
where $\sigma_{\min}$ is the smallest eigenvalue of the stationary state and $\lambda$ denotes 
the spectral gap of a particular symmetrization of the Liouvillian $\cL$. For reversible processes (semigroups satisfying detailed balance), $\lambda$ coincides with the spectral gap of $\cL$.  This convergence bound stems from a bound on the evolution of the quantum  
$\chi^2$-divergence. The $\chi^2$-divergence is defined as $\chi^2(\rho,\sigma) \equiv \tr{(\rho-\sigma)\sigma^{-1/2}(\rho-\sigma)\sigma^{-1/2}}$ and  yields an upper bound to the trace norm distance of the form $\|\rho-\sigma\|_{tr}^2\leq\chi^2(\rho,\sigma)$. As our example, let us consider the depolarizing semigroup on some Hilbert space $\cH \cong \bC^d$. The generator of a depolarizing semigroup acts on an observable $f$ as
\be
	\cL_{\rm depol}(f) =  \gamma\left(\frac{\1}{d}\tr{f}-f\right). 
\ee
The semigroup generated by $\cL_{\rm depol}$ is unital and has as its stationary state $\sigma = \1/d$. Furthermore, it satisfies detailed balanced and  thereby has a real spectrum \cite{chi2}. It is relatively easy to see, 
that the spectral gap $\lambda$ of $\cL_{\rm depol}$ is given by $\lambda = \gamma$. Furthermore, given that $\sigma = \1/d$, we get that 
$\sigma_{\min} = d^{-1}$. We are therefore left with the bound $\|\rho_t - \sigma \|_{tr} \leq \sqrt{d}e^{-\gamma t}$. Hence, we can give a bound on the mixing error by choosing $t \geq 1/\gamma \log\left (\sqrt{d}\epsilon^{-1}\right)$.  The mixing time bound derived from Eqn.~(\ref{chi2Mix}) then
scales as 
\be\label{chi2bound}
	\tau_{\chi^2} = {\cal O}( \log(d) ).
\ee

The trace norm allows for another upper bound, given in terms of the relative entropy   
$D\left(\rho \| \sigma \right) = \tr{\rho \left (\log(\rho) - \log(\sigma)\right)}$. Indeed, by the quantum Pinsker inequality \cite{PetzRel}, 
$\left\| \rho - \sigma \right \|_{tr}^2 \leq 2 D\left(\rho \| \sigma \right)$. In a spirit similar to the $\chi^2$-bound, we aim to give 
a bound on the evolution of the relative entropy. Assuming again, that $\rho_t$ evolves according to $\cL^*$, 
we find that the derivative of $D\left(\rho_t \| \sigma \right)$ is given by
\be
	\partial_t D\left(\rho_t \| \sigma \right) =  \tr{\cL^*(\rho_t)\left (\log(\rho_t) - \log(\sigma)\right)}.
\ee
The goal is to find a lower bound on the derivate of the relative entropy in terms of itself; i.e.
\be\label{ToyLS}
 2 \alpha_1 D\left(\rho_t \| \sigma \right) \leq - \tr{\cL^*(\rho_t)\left (\log(\rho_t) - \log(\sigma)\right)}.
\ee
Such a bound on the derivative of $D(\rho\|\sigma)$ leads to a time-dependent bound on the trace distance,
(c.f. theorem \ref{Mixingtimebound}), of the form
\be\label{ToyLSMix}
\left\| \rho_t - \sigma \right \|_{tr} \leq \sqrt{2\log\left(1/\sigma_{\min}\right)} \;\; e^{-\alpha_1 t}.
\ee
Note, the time-independent prefactor now only involves the logarithm of the smallest eigenvalue of $\sigma$. This can lead to a dramatic improvement of the mixing time bound, if the constant $\alpha_1$ is of the same order as the gap of the Liouvillian. We will show later, that 
the constant $\alpha_1$ is in fact always upper bounded by the spectral gap $\lambda$ for reversible Markov processes.

Returning to our example of the depolarizing channel $\cL_{\rm depol}$, we observe, that with $\sigma = \1/d$, we get
\be
- \tr{\cL^*_{\rm depol}(\rho_t)\left (\log(\rho_t) - \log(\sigma)\right)} =  \gamma D(\rho_t \| \sigma) + \gamma D(\sigma \| \rho_t) \geq \gamma D(\rho_t \| \sigma),
\ee
since $D(\sigma \| \rho) \geq 0$ for all states $\rho$. We are therefore led to the conclusion, that the inequality (\ref{ToyLS}) can be 
satisfied with the lower bound $\gamma/2 = \lambda/2 \leq \alpha_1$.  Given the bound (\ref{ToyLSMix}), we have to choose 
$ t \geq 2/\gamma\log(2\log(d)\epsilon^{-2})$ in order to ensure that the state $\rho_t$ deviates at most $\epsilon$ from the stationary state in trace distance. We therefore have that the Log-Sobolev bound (\ref{ToyLSMix}) gives an exponential improvement 
\be
	\tau_{LS} = {\cal O}(\log(\log(d))),
\ee
over the $\chi^2$- bound (\ref{chi2bound}) considered before.\\

The discussion we have given here illustrates the central idea of the logarithmic Sobolev (LS) inequality based approach to rapid mixing of 
continuous time Markov processes. The inequality given in Eqn.~(\ref{ToyLS}) is one example of a particular LS inequality. We will introduce the 
general framework shortly and also explain the connection it has to the phenomenon of hypercontractivity. 
The example of the depolarizing channel considered here is an instance, where the LS inequality approach gives an exponentially 
improved bound over the more common spectral gap approach. This will be true whenever the constant $\alpha_1$ (and subsequantly the spectral gap $\lambda$) is independent of the system size. More generally, whenever $\alpha_1$ and $\lambda$ are of the same order, the Log-Sobolev appraoch to mixing will be beneficial. To prove that this is the case is however a difficult task for specific problems, and in particular is not always the case. One example, where no improvement is found, is for expander maps. There, the LS constant scales with the system size in such a fashion that the improvement gained through the smaller pre-factor is rendered useless (see Sec. \ref{sec:Applications}).\\

\paragraph{\bf Informal exposition:}
The bound on the convergence we have stated in the example (Eqn. (\ref{ToyLSMix})), in terms of the relative entropy and the constant 
$\alpha_1$, is actually not the canonical Log-Sobolev inequality, which is commonly used to bound the mixing time in classical systems. 
The arguments which we have presented here rather correspond to the modified Log-Sobolev inequality (sometimes called entropy-entropy production inequalities) considered in \cite{Eep,ModLogSobolev}. 
In fact, there exists an entire family of Log-Sobolev inequalities indexed by some $p \in [1,\infty)$, 
which stem from arguments of hypercontractivity of the semigroup. To derive mixing time bounds, however, we only make use of two particular cases, which correspond to the values of $(p=1,2)$.  The inequality which corresponds to $p=2$, with constant $\alpha_2$, is the quantum generalization 
of the canonical classical inequality, which already has seen generalizations to the quantum setting in the aforementioned references \cite{Lieb,Gross72,Gross75,Ziggy1,Ziggy2}. This inequality for $\alpha_2$ does have the advantage of being in a simpler form than the rather involved 
inequality for $\alpha_1$. However, this simplicity comes at a price. It is not possible to immediately bound the trace norm in terms of the 
relative entropy and this constant ($\alpha_2$). For the $p=2$ Log-Sobolev inequality, further inequalities are necessary. In the quantum setting the derivation of these inequalities is hampered by the non-commutativity of the operators. In this paper we show that these inequalities hold for a large class of semigroups (referred to as $\bL_p$-regular). We devise a criterion (already considered in \cite{Ziggy1,Gross72,Gross75}) to verify $\bL_p$ regularity, which is related to the convexity of a trace-norm function.  We then show that some of the most commonly used semigroups satisfy $\bL_p$-regularity. We conjecture, that this condition in fact holds for all semigroups on finite state spaces.
The organization of the article is as follows: \\

In  {\it Section \ref{sec:Framework}} we define the {\it formal framework} of non-commutative $\bL_p$ spaces; originally 
introduced in \cite{Ziggy1,Ziggy2,LpSpaces1,LpSpaces2}. These spaces constitute the formal backbone of the hypercontractivity results for quantum semigroups. The central element is the $\bL_p$-norm $\|a\|_{\sigma,p}$, which is weighted with respect to some full rank density matrix $\sigma$.
In our analysis, it will be chosen to correspond to the stationary state of the semigroup.  We state some elementary results for these spaces 
and connect them to a class of relative entropy functionals. 

We then proceed to define a family of quantum Dirichlet forms $\cE_p(f)$ for the generator $\cL$ of the quantum semigroup, 
which will be the starting point for the definition of quantum logarithmic Sobolev inequalities.  A central result of this section 
is the definition of  {\it $\bL_p$-regularity}, which allows to relate different Dirichlet forms in this family to each other and  to 
prove a partial ordering  of Log-Sobolev constants relating $\alpha_2$ to the constant $\alpha_1$ used to derive 
the mixing time bounds.

In {\it Section \ref{HyperLSinq}} we define the general family of Log-Sobolev inequalities and establish the connection between 
{\it hypercontractivity and  LS inequalities}. A lower bound to the spectral gap $\lambda$ of a reversible 
generator $\cL$ in terms of the $LS_2$-constant $\alpha_2$ has already been proven in \cite{Ziggy1}. The main contribution
of this chapter is the lower bound on the spectral gap in terms of $\alpha_1$, which in turn implies the bound on $\alpha_2$ 
for $\bL_p$-regular channels due to the partial order of the Log-Sobolev constants.

We then investigate in more detail the $\bL_p$ regularity condition and its connection to a trace functional, which was already 
considered in \cite{Ziggy1}. We show that convexity of this trace functional implies the $\bL_p$-regularity condition. We then use this functional form to show that several important families of Liouvillians, including Davies generators \cite{Davis,Davis2}, satisfy $\bL_p$-regularity.

{\it Section \ref{sec:Mixing}} is devoted to the rigorous derivation of {\it mixing time bounds}. The results proved in the 
preceding sections are put together, and a formal derivation of the mixing time bounds is given. We point out a physical interpretation
of the $LS_1$ inequality and the associated quantities, and state the mixing time bounds, both in terms of $\alpha_1$ and $\alpha_2$.

Finally, in {\it Section \ref{sec:Applications}}, we consider {\it Applications} of the aforementioned results and 
derive mixing time bounds for some well-known simple generators. Here we also show, how the hypercontractivity 
of the associated semigroup can be used to obtain bounds on the Log-Sobolev constants. We provide a brief outlook in {\it Section \ref{sec:Conclusion}}.

\section{Formal framework}\label{sec:Framework}

To start with, we will need to introduce the necessary formal framework. Logarithmic Sobolev inequalities and their connection to hypercontractivity on quantum state spaces are most naturally formulated in the 
language of non-commutative  $\bL_p$  spaces, previously defined and analyzed in \cite{Ziggy1,Ziggy2,LpSpaces1,LpSpaces2}.  In an effort to make this paper as self-contained as possible, we will restate many of the main results on non-commutative  $\bL_p$ spaces, and introduce them in a self consistent manner. 

\bigskip

Throughout this paper we will be working exclusively with operators acting on finite Hilbert spaces ($d$-dimensional), which are isomorphic to the algebra of $d$-dimensional complex matrices $\cM_d \cong \bC^{d\times d}$, when equipped with an inner product. We denote the set of $d$-dimensional Hermitian operators $\cA_d=\{X\in\cM_d,X=X^\dag\}$, as well as the subset of positive definite operators $\cA^+_d=\{X \in \cA_d, X > 0\}$. The set of states will be denoted $\cS_d=\{X \in \cA_d,X\geq0,\tr{X}=1\}$, and the full rank states will be analogously denoted $\cS_d^+$. 
Observables will always be represented by lower case Latin letters ($f,g\in \cA_d$), and states by Greek letters ($\rho,\sigma \in \cS_d$). 

The central property of the non-commutative state spaces to be introduced below, is that the norm as well as the scaler product is weighted with respect to some 
full rank reference state $\sigma \in \cS_d^+$. This weighting can be expressed in terms of a map acting on elements  $f \in \cA_d$ by writing
\be
	\Gamma_\sigma(f) = \sigma^{1/2} f \sigma^{1/2}.
\ee

We would like to point out that this choice of $\Gamma_\sigma$ is not unique, in fact there exists an entire family of modular operators which could be used\footnote{The map $\Gamma_\sigma$ actually corresponds to the inverse of the modular operator present in Monotone Riemannian metrics of non-commutative information geometry \cite{chi2,RuskaiRiem}.}. These are  intimately related to monotone metrics on manifolds of quantum states; see \cite{chi2,Petz1,Petz2} and references therein for more details. This particular choice of $\Gamma_\sigma$ is however very natural in that it is itself a completely positive map and its particular form allows for simplified manipulations. For notational convenience we will also introduce powers 
of the operator $\Gamma_\sigma$ as $\Gamma_\sigma^p(f) = \sigma^{\frac{p}{2}} f \sigma^{\frac{p}{2}}$. The non-commutative $\bL_p$ spaces are equipped with a weighted  $\bL_p$-norm which, for any $f,g\in\cA_d$  and some $\sigma\in\cS_d^+$, is defined as
\be 
\| f \|_{p,\sigma} = \tr{\;|\; \Gamma_\sigma^{\frac{1}{p}}(f)\;|^p\;}^{\frac{1}{p}}=\tr{\;|\; \sigma^{\frac{1}{2p}} f \sigma^{\frac{1}{2p}}\;|^p\;}^{\frac{1}{p}}.
\ee
Similarly, the $\sigma$-weighted non-commutative $\bL_p$ inner product is given by 
\be
\avr{f,g}_\sigma = \tr{\Gamma_\sigma(f)g} = \tr{\sigma^{1/2} f \sigma^{1/2} g}.
\ee
Finally, we will also make extensive use of the $\bL_p$ variance which is defined as 
\be
\Var_\sigma(g) = \tr{\Gamma_\sigma(g) g} - \tr{\Gamma_\sigma(g)}^2. 
\ee 
It can easily be seen that, for any $g,f \in\cA_d$  and $\sigma\in\cS_d^+$, the variance is always positive ($\mbox{Var}_\sigma(g) \geq 0$), and that it is invariant under the transformation $g\raw g + c\1$, whenever $c\in\bR$.

In the remainder of the paper, unless specified otherwise, we will always be working with the $\bL_p$ norms and inner products. The reference state should always be clear from the context, and will almost always be the unique full rank stationary state of some Liouvillian.

In the following lemma, we summarize a number of important results concerning non-commutative $\bL_p$ spaces, which will be used repeatedly in the remainder of the paper. Proofs and discussions of these properties can be found in \cite{LpSpaces1,LpSpaces2}.

\begin{lemma}\label{Lem:Lp-norm}
The non-commutative $\bL_p$ spaces satisfy a:
\begin{enumerate}
\item  \textbf{Natural ordering of the $\bL_p$ norms:} Let $f\in \cA_d$ and $\sigma\in\cS_d^+$, then for any $p,q \in [1,\infty)$ satisfying $p\leq q$, we get $||f||_{p,\sigma}\leq||f||_{q,\sigma}$. 

\item \textbf{H\" older-type inequality:} Let $f,g\in \cA_d$ and $\sigma\in\cS_d^+$, then for any $p,q\in[0,\infty)$ satisfying $1/p+1/q=1$, 
\be |\avr{f,g}_\sigma|\leq ||f||_{p,\sigma}||g||_{q,\sigma}\ee

\item \textbf{Duality:} Let $f\in \cA_d$ and $\sigma\in\cS_d^+$, then for any $p,q\in[0,\infty)$ satisfying $1/p+1/q=1$, 
\be ||f||_{p,\sigma}=\sup\{\avr{g,f}_\sigma, g\in\cA_d, ||g||_{q,\sigma}\leq 1\}.\ee

\end{enumerate}
\end{lemma}

We now define several important functionals on the non-commutative $\bL_p$ spaces, and analyze their basic properties. These quantities are non-commutative $\bL_p$ versions of a number of known classical quantities. 

\begin{lemma}[The $\bL_p$ power operator]
Let $f\in\cA_d$, and $\sigma\in\cS_d^+$, then for any $p,q\in[1,\infty)$ define the \textbf{$\bL_p$ power operator} as:
\be 
I_{p,q}(f) =\Gamma_\sigma^{-1/p}\left[ |\Gamma_\sigma^{1/q}(f)|^{q/p}\right]= \sigma^{-\frac{1}{2p}}\left|\sigma^{\frac{1}{2q}} f \sigma^{\frac{1}{2q}}\right|^{q/p}\sigma^{-\frac{1}{2p}} 
\ee 
For any $f\in \cA_d$ and $p,q\in[1,\infty)$, it satisfies the following properties:
\begin{enumerate}
\item $||I_{p,q}(f)||^p_{p,\sigma}=||f||^q_{q,\sigma}$.
\item $I_{p,p}(f)=f$, and $I_{p,r}\circ I_{r,q}=I_{p,q}$, for any $r\in[1,\infty)$.
\item $I_{p,q}(cf)=c^{q/p}I_{p,q}(f)$, for any positive real $c\geq0$.
\end{enumerate}
\end{lemma}

The $\bL_p$ power operator acts in many ways like the usual matrix power operator. In particular, if the reference state is proportional to the identity ($\sigma=\1/d$), then $I_{p,q}(f)=f^{q/p}$. 

\bigskip

When acting on positive definite observables ($f\in\cA_d^+$), the infinitesimal structure of the non-commutative $\bL_p$ norms gives rise to an entropic functional which is intimately related to the relative entropy. Consider the directional derivative of the $\bL_p$ power operator  on the $\bL_p$ space, and define the \textbf{operator valued relative entropy} as

\be S_p(f)=-p\partial_s I_{p+s,p}(f)|_{s=0},\ee where for $f\in\cA_d^+$, and $s\geq0$, $S_p(f)$ can be evaluated explicitly and is given by
\be S_p(f)=\Gamma_\sigma^{-1/p}[\Gamma_\sigma^{1/p}(f)\log{[\Gamma_\sigma^{1/p}(f)]}]-\frac{1}{2p}\{ f,\log{\sigma}\}\ee

\begin{definition}[The $\bL_p$ relative entropy]\label{LpRelEnt}
Given $\sigma\in \cS^+_d$, and for any $f \in \cA^+_d$ we define the \textbf{$\bL_p$ relative entropy} to be
\be \Ent_p(f)=\avr{I_{q,p}(f),S_p(f)}_\sigma-||f||_{p,\sigma}^p\log{||f||_{p,\sigma}}\ee where $1/p+1/q=1$ and $p\geq1$.
\end{definition}

The $\bL_p$-regularized relative entropy and the $\bL_p$ norms can be further related by the following theorem, a proof of which can be found in \cite{Ziggy1}. 

\begin{theorem}\label{Thm:normRelEnt}
Given $f\in\cA_d^+$, we have that for any differentiable $p\equiv p(t)\geq1$,
\be \frac{d}{dt}||f||^p_{p,\sigma}=\dot{p}\avr{I_{q,p}(f),S_{p}(f)}_\sigma \label{Thm:normRelEntEq1} \ee 
where $1/p+1/q=1$. 
\end{theorem}

theorem \ref{Thm:normRelEnt} articulates the relationship that exists between 
the infinitesimal structure of $\bL_p$ norms and the $\bL_p$ relative entropy. This relationship is what 
enables the one-to-one correspondence between Log-Sobolev inequalities and hypercontractivity, 
which are, respectively, global and infinitesimal descriptions of the same contraction behavior of quantum dynamical semigroups.\\

Finally, we point out that the $\bL_p$ relative entropies with ($p=1,2$) play a special role within the family, and we will 
repeatedly make use of them in this paper. For the sake of clarity, we therefore write them out explicitly:

\begin{enumerate}
	\item{The $\bL_1$ relative entropy:} 
	\bq\label{ent1:def} \Ent_1(f) &=& \tr{\Gamma_\sigma(f)(\log(\Gamma_\sigma(f)) - \log(\sigma))} - \tr{\Gamma_\sigma(f)}\log(\tr{\Gamma_\sigma(f)}). \eq
	\item{The $\bL_2$ relative entropy:} 
	\bq \label{ent2:def} \Ent_2(f) &=& \tr{\left(\Gamma_\sigma^{1/2}(f)\right)^2 \log\left(\Gamma_\sigma^{1/2}(f)\right)} 
		 - \frac{1}{2} \tr{\left(\Gamma_\sigma^{1/2}(f)\right)^2 \log\left(\sigma\right)} \\ \nonumber
		    &&  -\frac{1}{2}\| f \|_{2,\sigma}^2\log\left(\| f \|_{2,\sigma}^2\right). \eq
\end{enumerate}

These two quantities can be related to each other and to the regular relative entropy as follows:

\begin{lemma}\label{RelEntProperties}
Let  $\sigma,\rho\in\cS_d^+$ and $f\in\cA_d^+$, then
\begin{enumerate}
\item $\Ent_2(I_{2,1}(f))=\half\Ent_1(f)$.
\item $\Ent_2(\Gamma_\sigma^{-1/2}(\sqrt{\rho}))= \frac{1}{2}D(\rho\|\sigma)$, where $D(\cdot\|\cdot)$ is the usual relative entropy.
\item $\Ent_1(\Gamma^{-1}_\sigma(\rho))=D(\rho\|\sigma)$
\item $\avr{I_{q,p}(f),S_{p}(f)}_\sigma = \frac{2}{p} \avr{I_{2,p}(f),S_{2}(I_{2,p}(f)}_\sigma$, for any $p,q\geq1$\\ satisfying $1/p+1/q=1$.
\end{enumerate}
\end{lemma}

\proof{
The above four identities can be obtained by straightforward manipulation of the quantities involved.\qed} 

\emph{Note:} it is clear that the $\bL_p$ relative entropies are ill-behaved for observables which are not strictly positive definite. In fact, as will be discussed later in Section \ref{sec:Mixing}, one can interpret the restriction to positive definite operators as a restriction to so called {\it relative densities}, which will be introduced later. These relative densities are the only type of operators which will be needed to derive the mixing time 
results.\\

\subsection{Dirichlet forms}
Throughout this paper, the time evolution of an observable ($f_t\in\cA_d$) will be described by one-parameter semigroups of completely positive trace preserving maps (cpt-maps), whose generator (Liouvillian) can always be written in standard \textit{Lindblad form} \cite{Lindblad}
\be
\partial_t f_t =  \cL(f_t) \equiv i[H,f_t] + \sum_i L^\dag_i f_t L_i - \frac{1}{2}\{L^\dagger_i L_i , f_t\}_+,
\ee
where $L_i \in  \cM_d$ are Lindblad operators and $H \in \cA_d$ is a Hamiltonian operator. We will denote the semigroup generated by $\cL$ by $T_t \equiv \exp(t\cL)$. This evolution corresponds to the dynamics in the Heisenberg picture, which specifies the dynamics on observables rather than states. We denote the dual of $\cL$, with respect to the Hilbert-Schmidt inner product, by $\cL^*$ which amounts to the evolution of states, i.e. the Schr\"odinger picture. The trace preserving condition ensures that $\cL(\1)=0$. If in addition $\cL^*(\1)= 0$, then the dynamics are said to be unital.

A Liouvillian $\cL$  is said to be \textit{primitive} if it has a unique full-rank stationary state. As the framework of non-commutative $\bL_p$ spaces depends on a full rank reference state, which will most often be the stationary state of a some dissipative dynamics, we will almost exclusively consider primitive Liouvillians. A discussion of primitivity in the context of quantum channels is given in \cite{Wielandt}, where several different characterizations are provided. 

A special class of Liouvillians which we will often consider are the ones which satisfy quantum detailed balance. A discussion about this class 
of maps and the corresponding conditions can be found for instance in \cite{chi2,db1,Majewski,Streater}. The definition we will be working with is the following:

\begin{definition}[Detailed balanced] We say a Liouvillian $\cL:\cM_d\rightarrow\cM_d$ satisfies \textbf{detailed balanced} (or is \textbf{reversible}) with respect to the state $\sigma\in\cS_d^+$, if $\Gamma_\sigma \circ \cL = \cL^* \circ \Gamma_\sigma$.
\end{definition}

The class of reversible generators has a number of particularly nice properties. The one most often exploited is that if $\cL$ satisfies detailed balance with respect to some $\sigma\in\cS_d^+$, then $\sigma$ is a stationary state of $\cL$.  Furthermore, the detailed balance condition ensures that the generator is Hermitian with respect to the weighted inner product $\avr{f,g}_\sigma$, which ensures that $\cL$ has a real spectrum. 
  
The particular Lindblad form of $\cL$ ensures that for any positive constant $t$, $e^{t\cL}$ is a cpt-map \cite{Lindblad}, which inherits the properties of unitality, primitivity and reversibility from its generator $\cL$. Unless otherwise specified, we will assume that the reference state of the non-commutative $\bL_p$ spaces is the unique full rank stationary state of some primitive Liouvillian, and we denote this stationary state $\sigma$; i.e. 
$\cL^*(\sigma)=0$. 

\bigskip

One of the fundamental tools in the classical theory of Log-Sobolev inequalities, and more generally in the classical theory of analytical methods for Markov chain mixing, is the Dirichlet form. We define a non-commutative $\bL_p$ regularized versions of it:

\begin{definition}[$\bL_p$ Dirichlet forms]
Given a primitive Liouvillian $\cL:\cM_d\rightarrow\cM_d$ with stationary state $\sigma$, we define its \textbf{$\bL_p$ Dirichlet form}:
\be \cE_p(f)=\frac{-p}{2(p-1)}\avr{I_{q,p}(f),\cL(f)}_\sigma, \ee for any $f\in\cA_d$, where $p\geq 1$ with $1/p+1/q=1$.
\end{definition}

The $\bL_p$ Dirichlet forms are  well defined even in the limit of $p=1$, which along with the $p=2$ case plays a special role in the remainder of the paper.  These two forms reduce to

\begin{proposition} The  $\bL_p$ Dirichlet forms for $p=1$ and $p=2$ are 
\begin{enumerate}
\item  For $p=2$,
\be \cE_2(f)= -\avr{f,\cL(f)}_\sigma. \ee
\item The limit $\lim_{p\raw1} \cE_p(f)$ exists an is given by
\be \cE_1(f)= -\frac{1}{2}\tr{\Gamma_\sigma(\cL(f))(\log(\Gamma_\sigma(f))-\log(\sigma))}\ee
\end{enumerate}
\end{proposition}

\proof{
Let $f\in\cA_d$, then $\cE_2(f)$ takes on this simple form by definition. When considering the form $\cE_1(f)$, 
observe that we have $\lim_{p\raw1}I_{p/(p-1),p}(f) = \1$, hence we can apply l'H\^opitales rule. We have 
\be\label{fugly1}
\lim_{p\raw1} \cE_p(f) = \left. -\frac{1}{2}\; \partial_p \;\avr{I_{p/(p-1),p}(f),\cL(f)}_\sigma\right |_{p=1},
\ee
So we need to compute $\partial_p I_{p/(p-1),p}(f) |_{p=1}$, for which one can see easily that
\be\label{fugly2}
\left. \partial_p I_{p/(p-1),p}(f) \right |_{p=1} = -\log(\sigma) + \log\left( \Gamma_\sigma(f) \right),
\ee
since we can write for the power operator
\be
I_{p/(p-1),p}(f) = \Gamma_\sigma^{(\frac{1}{p} - 1)} \left [
\exp\left({(p-1)\log\left( \Gamma_\sigma^{\frac{1}{p}} (f)  \right)}\right) \right ]. 
\ee
When we apply the product rule for the derivative and express 
\bq
\partial_p \; \left. \exp\left({(p-1)\log\left( \Gamma_\sigma^{\frac{1}{p}} (f)  \right)}\right)\right |_{p=1} &=& \partial_p \; \left. \sum_{n=0}^{\infty}\frac{(p-1)^{n}}{n!}\log\left( \Gamma_\sigma^{\frac{1}{p}} (f)  \right)^n\right |_{p=1}\\\nonumber &=& \log\left( \Gamma_\sigma(f)  \right), 
\eq we get the desired result. Inserting (\ref{fugly2}) into (\ref{fugly1}) we are left with the form $\cE_1(f)$ as stated in the proposition.\qed}

Note that, as its name suggests, the Dirichlet form usually has two distinct arguments $\cE(f,g)$. However, in all of the following we will consider these arguments to be identical, and hence we do not feel the need to define the more general form.  It was shown in \cite{Ziggy2} that, $\cE_p(f)\geq0$ for all $p\in[1,\infty)$ and for any $f\in\cA_d$. In the special case when $p=2$ then we additionally get that the Dirichlet form is invariant under the transformation $f\rightarrow f+c$. In the remainder of the paper, when referring simply to \textit{the} Dirichlet form, we mean the $\bL_2$ Dirichlet form.

In order to relate the Dirichlet forms for different $p$, in particular for $p=1$ and $p=2$, we need to introduce certain regularity conditions. These conditions will  play a crucial role in the remainder of the work.

\begin{definition}[$\bL_p$-regularity]\label{LpRegularity}
We say that the Liouvillian $\cL:\cM_d\rightarrow\cM_d$ is \textbf{weakly $\bL_p$-regular} if for all $p\geq1$, and all $f\in\cA_d$, we have
\bq \cE_p(f) \geq \left\{\begin{array}{rl} \cE_2(I_{2,p}(f)),~~&1\leq p\leq2\vspace{0.1cm} \\ 
(p-1)\cE_2(I_{2,p}(f)),~~& p\geq2.\end{array}\right.\eq\label{WeakLpReg}
Furthermore, we say that $\cL$ is \textbf{strongly $\bL_p$-regular} if for all $p\geq1$, and all $f\in\cA_d$, we have
\be \cE_p(f)\geq\frac{2}{p}\cE_2(I_{2,p}(f))\ee\label{StrongLpReg}
\end{definition}

A variant of strong $\bL_p$-regularity was already considered in \cite{Ziggy1}, where only reversible generators were investigated. 
Indeed, one can construct examples of non-reversible balanced channels which do not obey the strong $\bL_p$ regularity condition, by considering 
classical generators and embedding them into the present framework. The weak $\bL_p$-regularity condition is a generalization which allows to prove hypercontractivity and mixing time results, even when the channel is not reversible.  

As already pointed out, the $p=1$ and $p=2$ Dirichlet forms are the most relevant to us in our exposition and analysis. The $p=2$ form allows for simple access to the spectral gap of the Liouvillian via a variational characterization. Note that, in general when referring to the spectral gap of a Liouvillian, one mostly focuses on reversible maps, since otherwise the spectrum of the Liouvillian can be non-real. However, it is also possible in the general case to define a real constant $\lambda$ which relates\footnote{This constant would indeed be the spectral gap of the additive symmetrization of the Liouvillian: $\half (\cL+\Gamma_\sigma \cL^* \Gamma_\sigma^{-1})$.}  to the mixing time of the semigroup
in the same fashion as the gap does for reversible Liouvillians.  A more detailed discussion can be found in \cite{chi2}.  

\begin{definition} The \textbf{spectral gap} $\lambda$ of the the primitive Liouvillian $\cL:\cM_d\rightarrow\cM_d$ with stationary state $\sigma$ is defined as 
\bq \label{spectGapVar}
	\lambda  = \min \left \{ \left. \frac{\cE_2(g)}{\mbox{Var}_\sigma(g)} \Sp \right | \Sp g \in \cA_d, \;\; \Var_\sigma(g) \neq 0 \right\}.
\eq  
\end{definition}
One can easily verify that $\cE_2(g)$ is real and positive for all $g\in\cA_d$. This follows from the fact that the $\bL_2$ Dirichlet form of $\cL$ and of $\half (\cL+\Gamma_\sigma \cL^* \Gamma_\sigma^{-1})$ are equal.  Note, that the minimum is  actually attained by choosing $g$ as the eigenvector that corresponds to the first non vanishing eigenvalue, i.e. the spectral gap $\lambda$ of the Liouvillian symmetrization. We will see later that, as a consequence, this is also the relevant constant for non-reversible balanced maps when we bound the $\bL_2$ - mixing time.

\section{Hypercontractivity and Log-Sobolev inequalities}
\label{HyperLSinq}

In this section, we introduce the Log-Sobolev inequalities and prove their basic properties. In particular, we show that they are equivalent to hypercontractivity of the semigroup. We show that there exists a partial ordering of the Log-Sobolev constants for different $p$, and that the $p=1,2$ Log-Sobolev constants lower bound the spectral gap of the Liouvillian. 

Let us start by formally defining a set of general Log-Sobolev inequalities. We will later see that only two special cases ($p=1,2$) will be of interest to us, but it will often be convenient to work with the entire family.

\begin{definition}\label{Def:LSI}
Let $1/p+1/q=1$, with $q\geq1$, and let $\cL:\cM_d\rightarrow\cM_d$ be a Liouvillian. Assume that $\cL$ has a full rank stationary state ($\sigma\in\cA_d^+$). We say that $\cL$ satisfies a $p$-Log-Sobolev inequality ($LS_p$), if there exists a positive constant $\alpha_{p}>0$ such that
\be \alpha_p\Ent_p(f)\leq\cE_p(f),\label{Eqn:LSI}\ee for all $f\in\cA_d^+$. We call the largest $\alpha_p$ for which Eqn.~(\ref{Eqn:LSI}) 
holds the Log-Sobolev constant. 
\end{definition}

We will often simply say that "$LS_p$ holds" to mean that $\cL$ satisfies a $p$-Log-Sobolev inequality. It should be noted that this definition of the generalized Log-Sobolev inequalities reduces to the well-known classical definition given for instance in \cite{frencharticle}, when restricted to commutative state spaces. We will also need a working definition of hypercontractivity in order to state and prove the main theorems of this section. 

\begin{definition}\label{Def:Hyper}
Let $\cL:\cM_d\rightarrow\cM_d$ be a Liouvillian, and let $T_t$ be its associated semigroup. Assume that $\cL$ has a unique full rank stationary state ($\sigma\in\cA_d^+$). If 
\be ||T_t(f)||_{p(t),\sigma}\leq ||f||_{2,\sigma}\ee whenever $p(t)=1+e^{2\alpha t}$ for some $\alpha>0$, then the semigroup is said to be \textit{Hypercontractive}.
\end{definition}

We note that a slightly more general definition of hypercontractivity can be given where we define $p(t)=1+(p_0-1)e^{2\alpha t}$, and then consider the contraction of $||T_t(f)||_{p(t),\sigma}\leq||f||_{p_0,\sigma}$. However, as it is customary in the literature to consider $p_0=2$, and all of the applications considered in this paper use only the $p_0=2$ case, we chose it as our definition. 

Before proving the equivalence between Log-Sobolev inequalities and hypercontractivity, we will need to establish a partial ordering between the Log-Sobolev inequalities. The relation between the different $LS_p$ are summarized in the following proposition:

\begin{proposition}\label{LS2impliesLSq}
Let $\cL:\cM_d\rightarrow\cM_d$ be a primitive Liouvillian with stationary state $\sigma$. If  $\cL$ is strongly $\bL_p$-regular, then $\alpha_2\leq\alpha_p$ for all $p\geq1$. If  $\cL$ is weakly $\bL_p$-regular, then $\alpha_2\leq2\alpha_p$ for all $p\geq1$.
\end{proposition}

\proof{
This lemma follows by simple manipulation of the non-commutative $\bL_p$ norms and inner products. Indeed, by lemma \ref{RelEntProperties}
\be \avr{I_{q,p}(f),S_{p}(f)}_\sigma = \frac{2}{p} \avr{I_{2,p}(f),S_{2}(I_{2,p}(f)}_\sigma\ee if $1/p+1/q=1$. And since  $||I_{2,p}(f)||_{2,\sigma}^2=||f||_{p,\sigma}^p$ for all $f\in\cA_d$ and all $q\geq1$, then setting $g\equiv I_{2,p}(f)$,  $LS_2$ implies

\bq \frac{2}{p}\avr{g,S_2(g)}_\sigma -\frac{2}{p}||g||_{2,\sigma}^2\log{||g||_{2,\sigma}} &=& \avr{I_{q,p}(f),S_p(f))}_\sigma-||f||^p_{p,\sigma}\log{||f||_{p,\sigma}}\\ &\leq& \frac{2}{p \alpha_2}\cE_2(I_{2,p}(f))\eq

Hence, assuming strong $\bL_p$ regularity, we get $\Ent_p(f)\leq\frac{1}{\alpha_2}\cE_p(f)$ for all $f\in\cA^+_d$, while assuming weak $\bL_p$ regularity, we can only ensure that $\Ent_p(f)\leq\frac{2}{\alpha_2}\cE_p(f)$ for all $f\in\cA^+_d$.\qed}

This partial ordering will be very relevant when it comes to expressing mixing times bounds in terms of Log-Sobolev constants. 

\bigskip
In order to state the main theorem relating Log-Sobolev inequalities to hypercontractivity, we will need an essential lemma (first proved in \cite{Ziggy1} Sec. 3), which relates the $\bL_p$ norms to $LS_p$ inequalities.

\begin{lemma}\label{HyperLemma}
Let $\cL:\cM_d\rightarrow\cM_d$ be a primitive Liouvillian with stationary state $\sigma$,  and define $p\equiv p(t)\equiv1+e^{2\alpha t}$, with $\alpha>0$. Then, for $f\in\cA_d^+$, we have
\be \frac{d}{dt}\log{||f_t||_{p,\sigma}}=\frac{\dot{p}/p}{||f||_{p,\sigma}^p}\{\Ent_p(f)-\frac{1}{\alpha}\cE_p(f)\},\ee where $1/p+1/q=1$, and $p\geq2$.
\end{lemma}

 The proof is provided in \cite{Ziggy1}. The next theorem relates $LS_2$ to hypercontractivity of the semigroup and constitutes the cornerstone of the abstract theory of Log-Sobolev inequalities. 

\begin{theorem}\label{HypervsLS}
Let $\cL:\cM_d\rightarrow\cM_d$ be a primitive Liouvillian with stationary state $\sigma$, and let $T_t$ be its associated semigroup. Then
\begin{enumerate}
\item If there exists an $\alpha>0$ such that for any $t>0$, $||T_t(f)||_{p(t),\sigma}\leq||f||_{2,\sigma}$ for all $f\in\cA_d^+$ and $2\leq p(t)\leq1+e^{2\alpha t}$. Then $\cL$ satisfies $LS_2$ with $\alpha_2\geq\alpha$. 
\item If $\cL$ is weakly $\bL_p$-regular, and has an $LS_2$ constant $\alpha_2 >0$, then $||T_t(f)||_{p(t),\sigma}\leq||f||_{2,\sigma}$ for all $f\in\cA_d^+$, and any $t>0$ when $2\leq p(t)\leq1+e^{\alpha_2 t}$. If, furthermore, $\cL$ is strongly $\bL_p$ regular, then the above holds for all $t>0$ when $2\leq p(t)\leq1+e^{2\alpha_2 t}$.
\end{enumerate}
\end{theorem}
\proof{
We start by proving the first statement: "hypercontractivity implies Log-Sobolev inequality". The hypercontractivity condition and convexity of the logarithm imply that  for $p(t)=1+e^{2\alpha_2 t}$
\be \log{||f_t||_{p(t),\sigma}}<\log{||f||_{2,\sigma}} \ee for all $f\in\cA_d^+$.
Therefore, taking the derivative at $t=0$ from the right yields
\be \frac{d}{dt}||f_t||_{p(t),\sigma}|_{t=0}\leq 0\ee
Then using lemma \ref{HyperLemma}, we get that
\be \frac{d}{dt}||f_t||_{p(t),\sigma}|_{t=0} = \frac{\alpha}{||f||_{2,\sigma}^{2}}(\Ent_2(f)-\frac{1}{\alpha}\cE_2(f))\leq0\ee
which immediately implies $LS_{2}$ with $\alpha_2=\alpha$.

Now, for the inverse implication, assume that $\cL$ satisfies a Log-Sobolev inequality with $LS_2$ constant $\alpha_2$. If $\cL$ is weakly $\bL_p$ regular, then by proposition \ref{LS2impliesLSq}, $\cL$ satisfies a $LS_p$ with $2\alpha_p\geq\alpha_2$, for any $p\geq2$. In particular, lemma \ref{HyperLemma} guarantees that for $p(t)=1+e^{\alpha t}$,

\be \frac{d}{dt}||f_t||_{p(t),\sigma} \leq 0 \ee
Integrating this expression from $0$ to $t$ immediately gives hypercontractivity of the semigroup.
If, furthermore, $\cL$ is strongly $\bL_p$ regular, then the same reasoning guarantees hypercontractivity with $p(t)=1+e^{2\alpha t}$ \qed}

Hypercontractivity can be seen as a global statement of the contractivity of the semigroup, while the Log-Sobolev inequality is the equivalent infinitesimal statement. Depending upon the task at hand, it might be more convenient to work in one picture or the other. See, for instance, the analysis of expanders maps in Sec. \ref{sec:Applications} which crucially builds on this correspondence. 

Hypercontractivity provides a quantitative statement of the (worst case) convergence behavior of a map (semigroup) whereas simple contractivity just guarantees that the map is monotone. 
Hypercontractivity is a statement at the operator level (Heisenberg picture), and is hence much more amenable to infinite dimensional analysis. In fact the roots of the framework and of the tools introduced in this paper were developed for infinite dimensional systems \cite{Gross72,Gross75}. 

\bigskip

Finally,  we state the main new result of this section, which relates  the Log-Sobolev constant $\alpha_1$ to the spectral gap $\lambda$, for primitive reversible Liouvillians. A result relating the $LS_2$ constant $\alpha_2$ to twice the spectral gap was first proved in \cite{Ziggy1} for reversible Liouvillians. The following can be seen as a strengthening of their result for $\bL_p$-regular Liouvillians.

\begin{theorem} \label{Thm:LSgap} Let $\cL:\cM_d\rightarrow\cM_d$ be a primitive reversible Liouvillian with stationary state $\sigma$. 
The Log-Sobolev constant $\alpha_1$ and the spectral gap $\lambda$  of $\cL$ are related as: 
\be
	\alpha_1 \leq \lambda.
\ee
\end{theorem}

\Proof{ 
Let $g\in\cA_d$, and define $f_\epsilon=\1+\epsilon g$, where $\epsilon\in\bR^+$ is chosen in such a way that $f_\epsilon\in\cA_d^+$. Clearly, as $\epsilon\rightarrow0$, this is true. We now expand both sides of the inequality 
\be\label{ineqTheo1}
	 \alpha_1 \mbox{Ent}_1(f_\epsilon) \leq  \cE_1(f_\epsilon),
\ee 
in powers of $\epsilon$ up to second order. \\

Let us first focus on the left side of the inequality.  We have that
\be\label{leftInq}
\mbox{Ent}_1(f_\epsilon) =  \tr{\Gamma_\sigma(f_\epsilon)\left(\log(\Gamma_\sigma(f_\epsilon)) - \log(\sigma)\right)} 
		      - \tr{\Gamma_\sigma(f_\epsilon)}\log\left(\tr{\Gamma_\sigma((f_\epsilon)}\right)
\ee

We start by expanding the terms which involve $\tr{\Gamma_\sigma(f_\epsilon)}$. We immediately have that 
$\tr{\Gamma_\sigma(f_\epsilon)} = 1 + \epsilon \; \tr{\Gamma_\sigma(g)}$, which due to the Taylor expansion of the 
natural logarithm yields
\be
	\tr{\Gamma_\sigma(f_\epsilon)}\log\left(\tr{\Gamma_\sigma(f_\epsilon)}\right) = 
	\epsilon \; \tr{\Gamma_\sigma(g)} + \frac{\epsilon^2}{2} \tr{\Gamma_\sigma(g)}^2 + {\cal O}(\epsilon^3).
\ee

We now turn to the expansion of the remaining contributions in the renormalized entropy. 
For the expansion of the first term, we need to use the integral representation of the logarithm of an operator,

\be\label{log_ident}
	\log(A) = \int_0^\infty \frac{1}{t} - \frac{1}{t +  A} dt.
\ee
The difference $\log(A) - \log(B)$ can also be expressed as 
\be
	\log(A) - \log(B) = \int_0^\infty \frac{1}{t + B}\left(A   - B\right)\frac{1}{t +  A} dt,
\ee
due to the operator identity 
\be\label{opIdentity}
	A^{-1} - B^{-1} = A^{-1}\left(B -  A\right)B^{-1}.
\ee 
The expression  $\tr{\Gamma_\sigma(f_\epsilon)\left(\log(\Gamma_\sigma(f_\epsilon)) - \log(\sigma)\right)} $ can be written  
in terms of the following operator function
\be
	 \log(\Gamma_\sigma(f_\epsilon)) - \log(\sigma) = \epsilon \int_0^\infty \frac{1}{t + \sigma}\; 
	 \Gamma_\sigma(g) \; \frac{1}{t + \sigma + \epsilon \Gamma_\sigma(g)} dt,
\ee
which we will expand up to second order in $\epsilon$. Note that we can write
\bq
	\left( t + \sigma + \epsilon \Gamma_\sigma(g) \right)^{-1} &=&
	 (t + \sigma)^{-1} - (t + \sigma)^{-1}  +  \left( t + \sigma + \epsilon \Gamma_\sigma(g) \right)^{-1} \\\nonumber
	 &=& (t + \sigma)^{-1} - \epsilon (t + \sigma)^{-1} \Gamma_\sigma(g)  \left( t + \sigma + \epsilon \Gamma_\sigma(g) \right)^{-1},
\eq
due to the operator identity of Eqn. (\ref{opIdentity}). This Dyson like recursion for $\left( t + \sigma + \epsilon \Gamma_\sigma(g) \right)^{-1}$ gives rise to the following expansion
\bq\label{logExp2}
\log(\Gamma_\sigma(f_\epsilon)) - \log(\sigma)&& = \epsilon \int_0^\infty \frac{1}{t + \sigma}\; 
	 \Gamma_\sigma(g) \; \frac{1}{t + \sigma}\; dt \no - 
	 &&\epsilon^2  \int_0^\infty \frac{1}{t + \sigma}\; 
	 \Gamma_\sigma(g) \; \frac{1}{t + \sigma}\; \Gamma_\sigma(g) \; \frac{1}{t + \sigma}\; dt + {\cal O}(\epsilon^3).
\eq
Since $\Gamma_\sigma(f_\epsilon) = \sigma + \epsilon \Gamma_\sigma(g)$ we are left with the following approximation up to second order 
\bq
&&\tr{\Gamma_\sigma(f_\epsilon)\left(\log(\Gamma_\sigma(f_\epsilon)) - \log(\sigma)\right)}  = 
\epsilon \int_0^\infty \tr{\frac{\sigma}{(t + \sigma)^2} \Gamma_\sigma(g)}\; dt \no
&& - \epsilon^2\int_0^\infty \tr{\frac{\sigma}{(t + \sigma)^2} \Gamma_\sigma(g) \frac{1}{t + \sigma} \Gamma_\sigma(g)}  dt \no
&&+ \epsilon^2 \int_0^\infty \tr{\frac{1}{t + \sigma} \Gamma_\sigma(g) \frac{1}{t + \sigma} \Gamma_\sigma(g)} dt  + {\cal O}(\epsilon^3).
\eq
The integrals are conveniently evaluated in the basis in which $\sigma = \sum_\alpha \sigma_\alpha \proj{\alpha}$ is diagonal. It follows that
\be\label{stupid}
\int_0^\infty \tr{\frac{\sigma}{(t + \sigma)^2} \Gamma_\sigma(g)}\; dt = \tr{\Gamma_\sigma(g)}.
\ee
The other two integrals which occur at order $\epsilon^2$ evaluate to
\bq\label{badInt}
 \int_0^\infty \tr{\frac{\sigma}{(t + \sigma)^2} \Gamma_\sigma(g) \frac{1}{t + \sigma} \Gamma_\sigma(g)} dt
  = \frac{1}{2} \sum_{\alpha,\beta} \frac{\sigma_\alpha \sigma_\beta}{\sigma_\beta -  \sigma_\alpha }\log\left(\frac{ \sigma_\beta }{ \sigma_\alpha}\right) |\bra{\alpha}g\ket{\beta}|^2,
\eq
and
\bq\label{goodInt}
\int_0^\infty \tr{\frac{1}{t + \sigma} \Gamma_\sigma(g) \frac{1}{t + \sigma} \Gamma_\sigma(g)} dt 
 = \sum_{\alpha,\beta} \frac{\sigma_\alpha \sigma_\beta}{\sigma_\beta -  \sigma_\alpha }\log\left(\frac{ \sigma_\beta }{ \sigma_\alpha}\right) |\bra{\alpha}g\ket{\beta}|^2
\eq
We observe, that Eqn. (\ref{badInt}) is just $1/2$ of the integral in Eqn. (\ref{goodInt}). The expansion of the full renormalized entropy can therefore be expressed as 
\be
\mbox{Ent}_1(f_\epsilon) = \frac{\epsilon^2}{2}\left(  \int_0^\infty \tr{\frac{1}{t + \sigma} \Gamma_\sigma(g) \frac{1}{t + \sigma} \Gamma_\sigma(g)} dt 
- \tr{\Gamma_\sigma(g)}^2 \right) + {\cal O}(\epsilon^3).
\ee
The right side of Eqn. (\ref{ineqTheo1}) can also be expanded to second order in $\epsilon$  by making use of Eqn. (\ref{logExp2}) and by observing, that
$\Gamma_\sigma(\cL(f_\epsilon)) = \epsilon \Gamma_\sigma(\cL(g))$. We therefore have that
\be
{\cE}_1(f_\epsilon) =  - \frac{\epsilon^2}{2} 
\int_0^\infty \tr{\Gamma_\sigma(\cL(g))\frac{1}{t + \sigma} \Gamma_\sigma(g) \frac{1}{t + \sigma}} 
+ {\cal O}(\epsilon^3).
\ee
If we now divide both sides of Eqn. (\ref{ineqTheo1}) by $\epsilon^2 / 2$ and take the limit $\epsilon \raw 0$, we are left with
\be \label{almostPoincare}
 \alpha_1 \left(  \tr{\Gamma_\sigma(g) \; \Xi_\sigma(g)}  
- \tr{ \sigma \; g}^2 \right) \leq  -  \tr{\cL^*(\Gamma_\sigma(g)) \;  \Xi_\sigma(g)},
\ee
where we have defined the cpt-map 
\be\label{ximap}
\Xi_\sigma(A) = \int_0^\infty \frac{\sigma^{1/2}}{t + \sigma} A \frac{\sigma^{1/2}}{t + \sigma} dt. 
\ee 
This cpt-map is self-adjoined with respect to the canonical Hilbert-Schmidt scalar product and furthermore has the property that it commutes with the map 
$\Gamma_\sigma$. Direct computation in the eigenbasis of the stationary state $\sigma = \sum_\alpha \sigma_\alpha \proj{\alpha} > 0$ yields the spectrum of $\Xi_\sigma$, which is given by $\xi_{\alpha,\beta} = \frac{\sqrt{\sigma_\alpha \sigma_\beta }}{ \sigma_\beta - \sigma_\alpha} \log\left(\frac{\sigma_\beta}{\sigma_\alpha}\right)$. It can be verified easily that the spectrum obeys $0 < \xi_{\alpha,\beta} \leq 1$. Hence the map $\Xi_\sigma$ is a positive definite operator.

Let us now introduce new variables $v = \Gamma_\sigma^{1/2}(g)$ and define ${\cal Q}_\sigma = \Gamma_\sigma^{-1/2}\circ \cL^* \circ \Gamma_\sigma^{1/2}$. Eqn. (\ref{almostPoincare}) can now be rewritten as 
\be \label{almostPoincareII}
 \alpha_1 \left(  \tr{v \; \Xi_\sigma(v)}  - \tr{ \sigma^{1/2} \; v}\tr{ \sigma^{1/2} \; \Xi_\sigma(v)} \right) \leq  -  \tr{{\cal Q}_\sigma(v) \;  \Xi_\sigma(v)},
\ee
where we have made use of the fact that $\Xi_\sigma(\sigma^\alpha) = \sigma^\alpha$, for all $\alpha\in[0,1]$.  We change the notation 
for convenience. We denote by $\ket{v} = v \otimes \1 \ket{I}$ the vectorization of the matrix $v$ on ${\cal M}_d \cong \bC^{d^2}$, 
where $\ket{I} = \sum_k \ket{kk} \in \bC^{d^2}$. Furthermore, on this space the maps ${\cal Q}_\sigma$ and $\Xi_\sigma$ act as matrices, which
we denote by $Q$ and $S$ respectively. We can therefore rewrite Eqn. (\ref{almostPoincareII})
\be\label{almostPoincareIII}
	\bra{v} \left( \alpha_1\left(\proj{\sqrt{\sigma}} - \1 \right) - Q\right) S \ket{v} \geq 0.
\ee
If we define the matrix $L = \alpha_1\left(\proj{\sqrt{\sigma}} - \1 \right) - Q$ the problem of finding the lower bound to the gap $\lambda$ in terms 
of the  Log-Sobolev constant $\alpha_1$ reduces to proving the positivity of the matrix $L$. Since $\bra{g} L^\dag S \ket{g}\in \bR$, we have to show that the positivity of the map $M = \frac{1}{2}\left(S \;L  + L^\dag \;S\right)$, which holds due to the inequality (\ref{almostPoincareIII}) implies that $L \geq 0$. Since we are considering generators that satisfy detailed balanced, this implies that the matrix $Q$ is Hermitian and therefore so is $L$. The equation 
\be
	M = \frac{1}{2}\left(S \;L  + L\;S\right) \geq 0,
\ee 
is equivalent to the well studied Sylvester equation $AX - XB = Y$, which posses the unique solution $ X = \int_0^\infty \exp(-At)Y\exp(Bt)dt$,
if the spectra of $A$ and $B$ are disjoined and positive definite as well as negative definite respectively, c.f. \cite{Bhatia} theorem VII.2.3. 
If we identify now $A = S$, $B = -S$ as well as $X = L$ and $2M = Y$, we see that we can write
\be
	L = 2\int_0^\infty e^{-S t} M e^{-S t} dt.
\ee
Hence we have that $L$ is positive semi-definite since it can be expressed as the convex sum of matrices 
congruent to $M \geq 0$. Thus, we have that $\bra{v} L \ket{v} \geq 0$, which upon rearranging and back substitution yields
\be
\alpha_1 \left(  \tr{\Gamma_\sigma(g) \; g}  
- \tr{\Gamma_\sigma(g)}^2 \right) \leq  -  \tr{\Gamma_\sigma(g) \;  \cL(g)}. 	
\ee  
Thus we have found that the constant $\alpha_1$ is a lower bound to the spectral gap $\lambda$. \qed}

Given the partial ordering of the Log-Sobolev constants for weakly $\bL_p$ - regular generators, we note that this result,
in particular, implies that the same holds for $\alpha_2 \leq \lambda$.  Note that furthermore a general bound on $\lambda$ as 
defined in Definition \ref{spectGapVar} can be given without the assumption of reversibility, when the generator $\cL$ is unital.

\begin{corollary}
Let $\cL:\cM_d\rightarrow\cM_d$ be a primitive unital Liouvillian with $LS_1$ constant $\alpha_1$ and spectral gap $\lambda$. Then, $\alpha_1\leq\lambda$. 
\end{corollary} 
\Proof{ This follows directly from the fact that the map $\Xi$ as defined in Eqn. (\ref{ximap}) is the identity, so Eqn. (\ref{almostPoincare}) immediately yields the bound for the symmetrization of $\cL$. \qed}

\subsection{$\bL_p$-regularity}\label{Sec:LpReg}

In this section we discuss the conditions of strong and weak regularity, and provide three important classes of examples where these conditions can be proved to hold. To start with, we show that these conditions follow from the analytical properties of a particular trace functional, also defined in \cite{Ziggy1}, as follows:

\begin{lemma}\label{hFunc}
Let $\cL:\cM_d\rightarrow\cM_d$ be a primitive Liouvillian with stationary state $\sigma$, and associated semigroup $T_t = e^{t{\cL}}$. Given $g\in\cA_d^+$ and $t>0$, define the one-parameter trace functional
\be
	h(s) = \tr{\sigma^{s/4} g^{2-s} \sigma^{s/4} T_t \left( \sigma^{-s/4} g^{s} \sigma^{-s/4} \right)}
\ee
on the real interval $s \in [0,2]$. If for all $g \in\cA_d^+$ and $ t \geq 0$, $h(s)$ is convex for $s\in[0,2]$ , then $\cL$ is weakly $\bL_p$-regular. Furthermore, if for all $g \in\cA_d^+$ and $t>0$, $h(s)$ is symmetric about $s=1$, and completely monotone for $s\in[0,2]$ , then $\cL$ is strongly $\bL_p$-regular. 
\end{lemma} 

\proof{
In order to prove the theorem, we first note that the function $h(s)$ can be written in terms of the $\bL_p$ inner product as
\be h(s)=\avr{I_{2/(2-s),2}(f), T_t \circ I_{2/s,2}(f)}_\sigma, \ee
for the choice $g = \sigma^{1/4} f \sigma^{1/4}$, and $f\in\cA_d^+$. Then it can be seen that $h(0)=h(2)= ||f||_{2,\sigma}^2$ for any $t\geq0$. 
Assume that $h(s)$ is convex in $s$, we therefore get the two inequalities 
\bq
	h(s) &\leq& (1-s)||f||_{2,\sigma}^2 + s h(1), \Sp \mbox{for all} \Sp s \in [0,1] \label{beach1}  \\
	h(s) &\leq& (2-s)h(1) + (s-1) ||f||_{2,\sigma}^2, \Sp \mbox{for all} \Sp s \in [1,2]. \label{beach2}	
\eq 
These inequalities correspond to the two secants which can be drawn from $h(0)$ to $h(1)$, as well as from $h(1)$ to $h(2)$ respectively. 
We now relate $p$ to $s$ via $p = 2/(2-s)$. For the first inequality (Eqn. (\ref{beach1})) we have that $s \in [0,1]$, which implies $p \in [1,2]$.
Similarly, we have for Eqn. (\ref{beach2}) that $p \in [2,\infty)$.  Let us focus on the inequality in Eqn. (\ref{beach1}). It implies that
\bq
\avr{I_{p,2}(f), T_t \circ I_{p/(p-1),2}(f)}_\sigma \leq \left(1 - \frac{2(p-1)}{p}\right)\|f\|^{2}_{2,\sigma}, +  \frac{2(p-1)}{p}\avr{f, T_t (f)}_\sigma .
\eq
If we now rearrange both sides of the inequality and devide by $1/t$, we can take the limit $t \raw 0$,  which because
$\cL=\lim_{t\rightarrow0}\frac{1}{t}(T_t-id)$ yields 
\bq
-\avr{f,\cL (f)}_\sigma \leq  - \frac{p}{2(p-1)} \avr{I_{p,2}(f), \cL \circ I_{p/(p-1),2}(f)}_\sigma
\eq
The substitution $f = I_{2,p}(f')$ now yields $\cE_2(I_{2,p}(f')) \leq  \cE_p(f')$. The second inequality (Eqn. (\ref{WeakLpReg})) 
in the definition for weak $\bL_p$ regularity is obtained from Eqn. (\ref{beach2}) by the same arguments.\\
The proof for strong $\bL_p$-regularity is very similar. Assume that for $s\in[0,2]$, $h(s)$ is completely monotone and symmetric about $s=1$. Complete monotonicity implies that $h$ is differentiable and that $(-1)^n h^{(n)}(s)\geq0$ for any $s\in[0,2]$. Hence, if for some $s_0$, $(-1)^n h^{(n)}(s_0)\geq0$ for all $n$, then this holds for any $s\in[0,2]$ by Taylor expansion. Observe, furthermore, that since $h(s)$ is symmetric about $s=1$, it follows that  $(-1)^n h^{(n)}(1)=0$ for all $n$ odd. Thus, 
\bq h(s)&=&h(1)+(s-1)^2\sum_{n=1}^\infty \frac{h^{(2n)}(1)(s-1)^{2n}}{(2n)!}\\
&\leq&h(1)+(s-1)^2\sum_{n=1}^\infty \frac{h^{(2n)}(1)}{(2n)!}\\
&=& h(1)+(s-1)^2(h(2)-h(1))\\
&=& h(1)+(s-1)^2(||f||_{2,\sigma}^2-h(1))\label{hsequality}
\eq
The above inequality can be rewritten in terms of the $\bL_p$ inner product as
\be \avr{I_{q,2}(f),T_t\circ I_{p,2}(f)}_\sigma\leq \frac{4(p-1)}{p^2} \avr{f,T_t(f)}_\sigma+\left(1-\frac{4(p-1)}{p^2}\right)||f||_{2,\sigma}^2\ee 
where $p\equiv2/s$ and $1/p+1/q=1$. This inequality gives rise to the strong $\bL_p$-regularity, by similar arguments as for the weak $\bL_p$-regularity case. 
\qed}

We note that these conditions on $h(s)$ are sufficient, but not necessary for proving $\bL_p$-regularity. There could exist examples of semigroups which satisfy weak or strong $L_p$-regularity without $h(s)$ being convex or completely monotone, respectively. In fact, the natural question arises of whether there exist semigroups which are not $\bL_p$-regular? We have performed rudimentary numerical searches, and have not found any such examples, which leads us to state the following conjecture:

\begin{conjecture}\label{Lpconjecture}
Let $\cL:\cM_d\rightarrow\cM_d$ be a primitive Liouvillian. Then $\cL$ is weakly $\bL_p$-regular. If, furthermore, $\cL$ is reversible, then it is strongly $\bL_p$-regular. 
\end{conjecture} 

It should be pointed out, that strong $L_p$ regularity cannot hold for all non-reversible Liouvillians, as we have come up with simple numerical counterexamples. 

\bigskip

\textbf{EXAMPLES}
We will consider three examples of frequently encountered Liouvillians which satisfy weak and/or strong $\bL_p$ regularity: a) primitive unital Liouvillians, b) projection Liouvillians, c) thermal Liouvillians. 

\bigskip

{\bf a) Primitive Unital Liouvillians:}
Let $\cL:\cM_d\rightarrow\cM_d$ be a primitive unital Liouvillian, with associated semigroup $T_t$. For any given $t>0$, let $\{A_j(t)\}$ be the Kraus operators of $T_t$. Given some $g\in\cA_d^+$ and $t>0$, it follows that: 
\be 
h(s) = \tr{g^{2-s}T_t(g^s)}=\sum_j\tr{g^{2-s}A_j^\dag(t) g^{s}A_j(t)}
\ee 
We can now work in the eigenbasis of $g = \sum_k g_k \proj{k}$ and see that the individual summands can be written as 
\be
\tr{g^{2-s}A_j^\dag(t) g^{s}A_j(t)} =  \sum_{kl} g_k^2\left(\frac{g_l}{g_k}\right)^s |\bra{l}A_j(t)\ket{k}|^2.\label{eqnUnital}
\ee 
Hence, $h(s)$ can always be written as the sum of exponentials in $s$ with positive weights and is thus always convex in $s$, irrespective of the specific form of the $A_i(t)$. If, on top of being unital, the primitive Liouvillian $\cL$ is reversible, then it satisfies strong $\bL_p$ regularity. This follows from from lemma \ref{hFunc}, because $h(s)$ is symmetric about $s=1$ by the detailed balance condition, and hence all odd derivatives are zero at $s=1$. Furthermore, every even derivative is positive by Eqn. (\ref{eqnUnital}), which implies that primitive reversible unital semigroups are strongly $\bL_p$-regular.

\bigskip

{\bf b) Projection Liouvillians: }
A projection Liouvillians is one whose semigroup projects onto a given density matrix $\sigma$ starting from any initial state. The Liouvillians can be written explicitly as $\cL(f)=\gamma(\tr{f\sigma}\1 - f)$, and the associated semigroup as $T_t( f ) = (1-e^{-t\gamma})\tr{\sigma f} \1+e^{-t\gamma}f$. If we choose $\sigma\in\cS_d^+$, then the projection Liouvillian is clearly primitive and reversible. 
Given some $g \in \cA_d^+$ and $t>0$, $h(s)$ can be written explicitly as
\be 
h(s)=(1-e^{-t\gamma})\tr{\sigma^{s}g^{2-s}}\tr{\sigma^{1-s}g^s} + e^{-t\gamma}\tr{g^2}
\ee
Then, expanding in the eigenbasis of $\sigma=\sum_a \sigma_a \proj{a}$ and of $g=\sum_k g_k\proj{k}$ yields
\be
h(s)=(1-e^{-t\gamma })\sum_{abkl}\left(\frac{\sigma_a g_l}{\sigma_b g_k}\right)^s g_k^2\sigma_b|\avr{a|k}|^2|\avr{b|l}|^2+e^{-t\gamma}\tr{g^2}
\ee
Thus, we can again write $h(s)$, up to an additive constant, as a sum of exponentials in $s$ weighted by positive coefficients. It follows by the same arguments as in example 1 that the function is symmetric about $s=1$ and completely monotone, and consequently that primitive projective semigroups are strongly $\bL_p$-regular.

\bigskip

{\bf c) Thermal Liouvillians:}
We call thermal Liouvillians, the subclass of Liouvillians which describe the dissipative dynamics resulting as the weak (or singular) coupling limit of a system coupled to a large heat bath. These Liouvillians are often called Davies generators \cite{Davis,Davis2}. See \cite{Spohn} for a clear derivation and a discussion of when this canonical form can be assumed.   

Thermal Liouvillians can always be written as 
\be \cL_\beta=\cL_0+\sum_{k,\omega}\cL_{k,\omega}\ee The individual terms are given by
\bq \cL_0(f)&\equiv& i[H,f]-\half\sum_{k,\omega} \eta_k(\omega)\{S^\dag_k(\omega)S_k(\omega),f\}_+ \\
\cL_{k,\omega} (f)&\equiv& \eta_k(\omega)S^\dag_k(\omega)fS_k(\omega),\eq where $\omega$ are the so-called Bohr frequencies and the $k$ index  reflects the couplings to the environment. In particular, $k$ can always be chosen such that $k\leq d^2$. $\eta_k(\omega)$ are the Fourier coefficients of the two point correlation functions of the environment, and are bounded. The $S_k(\omega)$ operators can be understood as mapping eigenvectors of $H$ with energy $\omega$ to eigenvectors of $H$ with energy $E+\omega$, and hence act in the Liouvillian picture as quantum jumps which transfer energy $\omega$ from the system to the bath. Thermal Liouvillians always satisfy detailed balance. In physical terms this means that it is as likely for the system to transfer an amount $\omega$ of energy to the environment as it is for the environment to transfer the same amount back to the system.

The thermal map can be seen to have a unique (full-ranked) stationary state which is given by $\sigma_\beta\propto e^{-\beta H}$, where $\beta$ is the inverse temperature of the heat bath. The following useful relations hold for any $k$ and $\omega$: 
\bq \eta_k(-\omega)&=&e^{-\beta \omega}\eta_k(\omega)\label{DBDavies1}\\ \sigma_\beta S_k(\omega)&=&e^{\beta \omega}S_k(\omega)\sigma_\beta \label{DBDavies2},\eq where Eqns. (\ref{DBDavies1}) and (\ref{DBDavies2}) are equivalent to the detailed balance condition for $\cL_\beta$. These in particular allow us to show the following:

\begin{theorem} 
Let $\cL_\beta:\cM_d\rightarrow\cM_d$ be the generator of a (bounded) thermal semigroup of $H$ at temperature $\beta$. Then $\cL_\beta$ is strongly $\bL_p$-regular.
\end{theorem}

\proof{
As in examples 1) and 2), we will show that for any $g\in\cA_d^+$ and $t>0$, the function $h(s)$ is completely monotone for $s\in[0,2]$. We point out that $h(s)$ can be rewritten as $h(s)=\tr{g^{2-s} \Gamma^{s/2}_{\sigma_\beta} \circ T_t\circ\Gamma^{-s/2}_{\sigma_\beta} (g^{s})}$, where $T_t\equiv e^{t\cL_\beta}$.

Observe that Eqns. (\ref{DBDavies1}) and (\ref{DBDavies2}) imply,
\be \Gamma^{s/2}_{\sigma_\beta} \circ \cL_\beta \circ\Gamma^{-s/2}_{\sigma_\beta}= \cL_0 +\sum_{k,\omega}e^{-s\beta\omega}\cL_{k,\omega} \label{mainIDThermal}\ee
Now, we want to invoke the quantum trajectories expansion of thermal semigroups (see \cite{Maes} for a recent account of this technique). It can be seen that a thermal semigroup can always be written as\footnote{This representation is not restricted to thermal semigroups, but in fact holds for any quantum dynamical semigroup. It appears under different names in the literature: quantum trajectories, unraveling of the master equation, quantum stochastic (or It\^o) calculus.}
\be \label{trajExpans}e^{t\cL}=\int d\xi \cW_t(\xi)\ee where writing $\Lambda_t\equiv e^{t\cL_0}$, 
\be \cW_t(\xi) = \Lambda_{t_1} \cL_{k_1,\omega_1}\Lambda_{t_2-t_1} ... \Lambda_{t_n-t_{n-1}}\cL_{k_n,\omega_n}\Lambda_{t-t_n},\label{EqnW}\ee and $\xi=(t_1,k_1,\omega_1; ...;t_n,k_n,\omega_n)$ is a given quantum trajectory with $0\leq t_1\leq ... \leq t_n\leq t$. The integral is shorthand for
\be \int d\xi \equiv \sum_{n,\vec{k},\vec{\omega}}\int \vec{dt}\equiv \sum_{n=0}^\infty \sum_{k_1,...,k_n}\sum_{\omega_1,...,\omega_n}\int_0^tdt_n\int_0^{t_n}dt_{n-1}...\int_0^{t_2}dt_1.\ee  

This expansion in terms of quantum trajectories is norm convergent if the operators  $\Lambda_t$ and $\cL_{k_n,\omega_n}$
have a bounded interaction rate, (c.f.  \cite{Davis}, ch. 5), that is if we have that for any such operator there exists a constant $K$
such that $\tr{\cW_t(\xi)(\rho)} \leq K \tr{\rho}$. Since we are working on a finite dimensional Hilbert space, this is clearly the case.

From Eqns. (\ref{mainIDThermal}) and (\ref{EqnW}), we get
\be \Gamma^{s/2}_{\sigma_\beta} \circ e^{t\cL}\circ \Gamma^{-s/2}_{\sigma_\beta} =\int d\xi e^{-s\beta\sum_n\omega_n}\cW_t(\xi)\ee

Finally, plugging this expression into $h(s)$, and considering the spectral decomposition $g=\sum_k g_k \proj{k}$, we get

\bq h(s)&=&\int d\xi e^{-s\beta\sum_n\omega_n} \tr{g^{1-s}\cW_t(\xi)(g^s)} \\ \label{serisConv}
&=& \int d\xi e^{-s\beta\sum_n\omega_n} \sum_{i,j}g_i^{1-s}g_j^s\tr{\ket{i}\bra{i}\cW_t(\xi)(\ket{j}\bra{j})} \\
&=& \sum_{n,\vec{\omega},i,j} e^{-s\beta\sum_n\omega_n} g_i^{1-s}g_j^s \int \vec{dt} \sum_{\vec{k}}\tr{\ket{i}\bra{i}\cW_t(\xi)(\ket{j}\bra{j})}\label{hstraj}\eq
$\cW_t(\xi)$ is a  completely positive map for any trajectory $\xi$, thus it allows for a Kraus decomposition. This implies that the trace in Eqn. (\ref{hstraj}) is always positive.  This, together with the convergence of (\ref{trajExpans}), ensures that (\ref{serisConv}) converges absolutely 
so we can exchange the summation and the integrals. We can now invoke the same arguments as in example 1) to justify that $h(s)$ is completely monotone and symmetric about $s=1$. Hence, by lemma \ref{hFunc}, we get that thermal Liouvillians are strongly $\bL_p$-regular.
\qed}

\bigskip

As we have already pointed out, we do not know of any Liouvillians which do not satisfy the $\bL_p$ regularity condition. This has led us to conjecture 
that this condition should always hold in finite dimensions. Proving this remains an open challenge. We have investigated three example classes of 
primitive Markov processes which we expect to find broad applications in quantum information theory.  In Sec. \ref{sec:Applications}, we will discuss a few of them. To the best of our knowledge, these three example classes also correspond to the only family of processes to which classical Log-Sobolev inequalities have been applied. The known applications of classical Log-Sobolev inequalities are processes which either converge to a uniform 
mixture (i.e. unital processes), for instance expanders and random walks, or to thermal processes which converge to the Gibbs distribution, such as the Glauber dynamics for the Ising model.

\section{Mixing time bounds}\label{sec:Mixing}

We are now ready to discuss the convergence behavior of dynamical semigroups on finite state spaces. To quantify the convergence behavior of 
these processes we need to choose an appropriate norm that quantifies the deviation from the stationary state of the process. The convergence is most
often estimated in trace norm, $\| A \|_{tr} = \tr{|A|}$, because of its operational interpretation \cite{Fuchs}, as being the optimal distinguishability between two states, when given access to arbitrary measurements. However, it is much more convenient to work with other distance measures, when considering mixing time bounds. The two more relevant measures we will work with are the $\chi^2$- divergence and the relative entropy. 
For any pair of states $\rho \in \cS_d$, and $\sigma \in \cS_d^+$, the following bounds are known \cite{chi2,PetzRel}

\bq\label{TraceBounds}
\left\| \rho - \sigma \right \|_{tr}^2  \leq \left \{ \begin{array}{l} \Sp \chi^2(\rho,\sigma) = \tr{(\rho-\sigma)\Gamma_\sigma^{-1}(\rho-\sigma)}
 \vspace{0.2cm} \\ \Sp  2\,D(\rho \| \sigma) = 2\, \tr{\rho\left(\log(\rho) - \log(\sigma)\right)} . \end{array} \right.
\eq

We have already stated in the introduction, that one of the main motivations for introducing Log-Sobolev inequalities 
in the finite system setting is to derive improved bounds on the convergence time.  The intuitive way of understanding the Log-Sobolev 
based bounds is by realizing, that the mixing time bound which is associated to the spectral gap $\lambda$ of the generator 
arises through bounding the dynamical behavior of the $\chi^2$-divergence, whereas the Log-Sobolev constant arises when 
one bounds the dynamics of the relative entropy directly. 

To establish the connection between trace norm mixing time bounds and the theory of Log Sobolev inequalities, let us introduce
the {\it relative density} of some state $\rho\in\cS_d$ with respect to the full rank state $\sigma\in\cS_d^+$,  which is defined as
\be 
\rho^\sigma \equiv  \Gamma_\sigma^{-1}(\rho) =\sigma^{-1/2} \rho \sigma^{-1/2}.
\ee 
This immediately allows us to relate the weighted 2-norm and the variance to the quantum $\chi^2$ - divergence. 
We have that  

\be
\chi^2(\rho,\sigma) = \tr{(\rho-\sigma)\Gamma_\sigma^{-1}(\rho-\sigma)} = \| \rho^\sigma -  \1 \|_{2,\sigma}^{2}=\Var_\sigma(\rho^\sigma). 
\ee

In the same way as for observables and states, we can define a natural dynamical equation for the relative density. It is given by the map

\be\label{compositeMAP}
	\hat{\cL} =  \Gamma_\sigma^{-1} \circ \cL^* \circ \Gamma_\sigma,  
\ee
which can be seen as the dual of $\cL$ with respect to the $\sigma$ weighted inner product $\avr{\cdot,\cdot}_\sigma$. This map defines the dynamics of the relative density via $\partial_t \rho^\sigma_t = \hat{\cL}(\rho^\sigma_t)$, as can easily be verified. Furthermore, it is again  a valid generator of a completely positive semigroup in the Heisenberg picture, with the same stationary state $\sigma$; i.e.  $\hat{\cL}(\1) = 0$ as well as $\hat{\cL}^*(\sigma) = 0$. The fact that $\hat{\cL}$ is the generator of a cpt-map follows from the particular form of $\Gamma_\sigma$, since  $\Gamma_\sigma$ as well as its inverse are completely positive maps. Hence we have that the composition $\exp( t \hat{\cL}) = \Gamma_\sigma^{-1} \circ \exp(t \cL^*) \circ \Gamma_\sigma$ is again completely positive. In general, when we want to construct mixing time bounds on the trace distance in terms of the Log-Sobolev inequalities, we will need to work with $\hat{\cL}$ instead of $\cL$. However, in the special  case where $\cL$ is \emph{reversible}, we have that $\hat{\cL} = \cL$. In order to avoid confusion, we will denote the corresponding Dirichlet forms, where we have replaced the generator $\cL$  with $\hat{\cL}$ by $\hat{\cE}_p$. Note however that for $p=2$, we get $\cE_2(f) =  \hat{\cE}_2(f)$ for every $f\in\cA_d$. 

We now proceed to derive the general mixing time bounds
  
\begin{lemma}\label{varLogBound}
Let $\cL:\cM_d\rightarrow\cM_d$ be a primitive Liouvillian with stationary state $\sigma$. 
\begin{enumerate}
\item Let $\lambda$ be the spectral gap of $\cL$; i.e. $\lambda \Var_\sigma(g) \leq \hat{\cE}_2(g)$ 
 for all $g \in \cA_d$. Then,
\be\label{VarianceBound}
\Var_\sigma(f_t) \leq e^{-2\lambda t} \Var_\sigma(f).
\ee
\item Let $\alpha_1$ be the $LS_1$ constant of $\hat{\cL}$; i.e.
$\alpha_1 \Ent_1(f) \leq \hat{\cE}_1(f)$, for all $f \in \cA_d^{+}$. Then,
\be\label{EntropyBound}
\Ent_1(f_t) \leq e^{ -2\alpha_1 t}  \Ent_1(f).
\ee
\end{enumerate}
\end{lemma}

\Proof{ 

1. We first bound the dynamical evolution of the variance in Eqn. (\ref{VarianceBound}). We find by simple calculation that $\partial_t \mbox{Var}_\sigma(g_t) =  - 2\hat{\cE}_2(g_t)$. The inequality for the gap $\lambda$ leads to the differential inequality $ \partial_t \mbox{Var}_\sigma(g_t) \leq -2\lambda\mbox{Var}_\sigma(g_t)$, which upon integration gives the desired bound.

2. The derivation of Eqn. (\ref{EntropyBound}) follows by similar arguments, we have to show that $\partial_t \Ent_1(f_t) = - 2\hat{\cE}_1(f)$ is obtained
as the derivative of the entropy functional. To see this, first note that $\partial_t \tr{\Gamma_\sigma(f_t)} = \tr{\cL(\Gamma_\sigma(f_t))} = 0$, since $\cL^*(\1) = 0$.
Hence, we obtain for the derivative
\be
\frac{d}{dt} \Ent_1(f_t) = \tr{\Gamma_\sigma(\hat{\cL}(f_t))(\log(\Gamma_\sigma(f_t))-\log(\sigma))} + \tr{\Gamma_\sigma(f_t)\left( \frac{d}{dt} \log(\Gamma_\sigma(f_t))\right)}
\ee 
It is possible to write the logarithm of a matrix by making use of the integral representation $\log(A) = \int_0^{\infty} \frac{1}{\lambda} - \frac{1}{\lambda + A} \; d\lambda $. It follows that,
\be
\frac{d}{dt} \log(\Gamma_\sigma(f_t)) = \int_0^\infty \frac{1}{\lambda + \Gamma_\sigma(f_t)} \left(\frac{d}{dt} \Gamma_\sigma(f_t) \right)  \frac{1}{\lambda + \Gamma_\sigma(f_t)} \;d\lambda.
\ee
Hence, we obtain
\bq
\tr{\Gamma_\sigma(f_t)\left(\frac{d}{dt} \log(\Gamma_\sigma(f_t))\right)} &=& \int_0^\infty \tr{\frac{\Gamma_\sigma(f_t)}{(\lambda + \Gamma_\sigma(f_t))^{2}} \left(\frac{d}{dt} \Gamma_\sigma(f_t) \right)} \;d\lambda \\ \nonumber &=& \frac{d}{dt} \tr{\Gamma_\sigma(f_t)} = 0,
\eq
which follows by direct integration in the eigenbasis of $\Gamma_\sigma(f_t)$. By comparison with the definition of $\hat{\cE}_1(f_t)$, we therefore have 
that $\partial_t \Ent_1(f_t) = -2\hat{\cE}_1(f_t)$. This yields, due to the Log-Sobolev inequality with constant $\alpha_1$, the desired bound by 
integrating the differential inequality $\partial_t \Ent_1(f_t) \leq - 2\alpha_1 \Ent_1(f_t)$. \qed}
 
This lemma implies the desired mixing time bounds for the trace norm distance via the relative entropy and the $\chi^2$-divergence.  Also note, that we did not make use of the fact that $\lambda$ corresponds to the second largest eigenvalue of $\cL$.  This fact  won't be true in general and only holds for reversible channels as we have already mentioned earlier. We have that in general that the spectrum of $\cL$ may actually be complex. However,  it can be seen that $\lambda$ can always be understood as the second largest eigenvalue of an appropriately weighted symmetrization \cite{chi2}. 

\begin{theorem}\label{Mixingtimebound}
Let $\cL:\cM_d\rightarrow\cM_d$ be a primitive Liouvillian with stationary state $\sigma$. Then the following trace norm
convergence bounds hold
\begin{enumerate}

\item{\bf $\chi^2$ bound}: If the inequality $\lambda \mbox{Var}_\sigma(g) \leq \hat{\cE}_2(g)$ holds for all $g \in \cA_d$, 
we have
\be
	\left\| \rho_t - \sigma \right \|_{tr} \leq \sqrt{1/\sigma_{\min}}e^{-\lambda t}.
\ee  

\item {\bf Log-Sobolev bound}: Furthermore if the $LS_1$ inequality 
$\alpha_1 \mbox{Ent}_1(f) \leq \hat{\cE}_1(f)$ holds for all $f \in \cA_d^+$, then
\be
	\left\| \rho_t - \sigma \right \|_{tr} \leq \sqrt{2\log(1/\sigma_{\min})}e^{-\alpha_1 t},
\ee
\end{enumerate} 
Where $\sigma_{\min}$ denotes the smallest eigenvalue of the stationary state $\sigma$.
\end{theorem}

\proof{ The theorem is a direct consequence of lemma \ref{varLogBound} and the bounds on the trace norm, in terms of the relative 
entropy and the $\chi^2$-divergence. 
Recall that, (c.f. lemma \ref{RelEntProperties}), we have $\mbox{Ent}_1( \Gamma_\sigma^{-1}(\rho)) = D(\rho \| \sigma)$ and $\mbox{Var}_\sigma( \Gamma_\sigma^{-1}(\rho)) = \chi^2(\rho,\sigma)$. Hence, lemma \ref{varLogBound} implies that 
$D(\rho_t \| \sigma) \leq e^{ - 2 \alpha_1 t}  D(\rho_0 \| \sigma)$ as well as\\
$\chi^2(\rho_t,\sigma) \leq e^{-2\lambda t} \chi^2(\rho_0,\sigma)$. With the bounds in Eqn. (\ref{TraceBounds}), we now have
\be
\left\| \rho_t - \sigma \right \|_{tr}  \leq \left \{ \begin{array}{l} \Sp \sqrt{\chi^2(\rho_0,\sigma)} \; e^{-\lambda t}
\vspace{0.2cm} \\ \Sp  \sqrt{2\,D(\rho_0 \| \sigma)} \; e^{- \alpha_1 t} . \end{array} \right.	
\ee 
Observe that both $D(\rho_0 \| \sigma)$ and $\chi^2(\rho_0,\sigma)$ become maximal for a full rank state $\sigma$, if $\rho_0$ corresponds 
to a rank one projector onto the eigenstate of $\sigma$ with the smallest eigenvalue, which leaves us with the stated bounds. \qed}

Note, that we have at no point made use of the fact that the generator of the semigroup is reversible or even $\bL_p$ - regular. The above 
results hold in general without further conditions on the Liouvillian $\cL$ apart from the assumption of primitivity, which ensures that the stationary state 
$\sigma$ has full rank. We have already mentioned in the introduction, that the logarithmic Sobolev inequality which corresponds to $\alpha_1$ 
is not the common inequality considered in the majority of classical mixing time results \cite{Diaconis,Martinelli,Martinelli2}, but was only somewhat recently introduced \cite{ModLogSobolev} to derive mixing time bounds for classical finite Markovian processes.\\

The partial ordering obeyed by the Log-Sobolev constants of $\bL_p$-regular Liouvillians was articulated in proposition \ref{LS2impliesLSq}; in particular, $\alpha_2 \leq 2\alpha_1$. In many practical applications, it seems to be more challenging to find lower bounds to the constant $\alpha_1$ than to the constant $\alpha_2$, 
since the Dirichlet form $\hat{\cE}_1(f)$ is more complicated than the standard Dirichlet form $\hat{\cE}_2(f) $.   This is why for classical 
Markov processes mostly the $LS_2$ inequality is used to derive mixing time bounds. For the class of $\bL_p$ regular generators
we are in fact able to reproduce the well-known classical result \cite{Diaconis} with the constant $\alpha_2$. 

\begin{lemma}
If the Liouvillian $\cL:\cM_d\rightarrow\cM_d$ is weakly $\bL_p$-regular and the $LS_2$ inequality $\alpha_2 \mbox{Ent}_2(f) \leq \hat{\cE}_2(f)$ holds for all 
$f \in \cA_d^+$, we have that
\be
 \left\| \rho_t - \sigma \right \|_{tr} \leq \sqrt{2\log\left(1/\sigma_{\min}\right)} \;\; e^{-\alpha_2 t/2}.
\ee 
Furthermore, if the Liouvilian $\cL$ is strongly $\bL_p$ regular, the bound can be improved to
 \be
 \left\| \rho_t - \sigma \right \|_{tr} \leq \sqrt{2\log\left(1/\sigma_{\min}\right)} \;\; e^{- \alpha_2 t}.
 \ee
\end{lemma}

\proof{
This bound follows immediately from theorem \ref{Mixingtimebound} and the partial ordering of the Log-Sobolev inequalities in proposition 
\ref{LS2impliesLSq}. \qed}

{\bf Remark 1:} We can provide good estimates of $\sigma_{\min}$ for the two situations which are of particular interest to us: primitive unital semigroups, and thermal semigroups. For primitive unital semigroups of a $d$-dimensional system, $\sigma=\1/d$, and hence $1/\sigma_{\min}=d$. For thermal semigroups of an $N$-qubit system with Hamiltonian $H$ at temperature $\beta$, the stationary state will be given by $\sigma_\beta=e^{-\beta H}/\tr{e^{-\beta H}}$. It is a straightforward calculation to see that we have the bound 
\be
 \frac{1}{\sigma_{\min}}\leq de^{\beta||H||_\infty}
\ee
Provided that the Hamiltonian is locally bounded and has only a polynomial number of terms, we get that for some positive constant $c\in\bR$, 
$\sigma_{\min}^{-1}\leq d e^{cN}$. For an $N$-qubit system we have that $d = 2^N$, which yields a scaling of $ \sigma_{\min}^{-1} \leq \cO(d)$.

This implies that for both of our cases of interest, the pre-factor in the Log-Sobolev bound grows at most as $\log(d)$. Hence, its contribution the the mixing time is of the order of $\log(\log(d))$. This indicates that the Log-Sobolev constant gives a very strong estimate on the mixing time. 

{\bf Remark 2:} The $LS_1$ inequality has the nice property that it allows for a convenient physical interpretation in terms of non-equilibrium 
thermodynamics. Consider a thermal Liouvillian $\cL_\beta$ with stationary state $\sigma_\beta = Z^{-1} \exp(-\beta H)$. If we compute the relative 
entropy of the evolved state $\rho_t$ with respect to the stationary state $\sigma_\beta$, we get
\be
	D(\rho_t \| \sigma_\beta) = \beta \left(H_t - TS_t  - F_\beta \right),
\ee
where we have denoted the Helmholtz Free-energy by $F_\beta = -k_B T \log(Z)$ with $\beta^{-1} = k_B T$  and the 
energy as well as the von Neumann entropy by $H_t = \tr{H\rho_t}$ and $S_t = -k_B\tr{\rho_t\log(\rho_t)}$, respectively.
For a thermal stationary state, the relative entropy is nothing but the difference between the thermal Free-energy and the 
non-equilibrium Free-energy. Furthermore, upon recalling that  $\rho^\sigma_t = \Gamma_\sigma^{-1}(\rho_t)$, the $LS_1$ 
Dirichlet form $\cE_1(\rho^\sigma_t)$ corresponds to the entropy production rate $\Pi$  of non-equilibrium Thermodynamics \cite{Spohn,Groot}. 
\be\label{entropyProduction}
        2 k_B \cE_1(\rho^\sigma_t) \equiv  \Pi = \frac{d}{dt}S_t + \Phi, 
\ee  
Here,  $\Phi = k_B \tr{\cL(\rho_t)\log(\sigma)}$ is often referred to as the entropy flux and we have that entropy production rate is given by 
$\frac{d}{dt}S_t = -k_B \tr{\cL(\rho_t)\log(\rho_t)}$.  The rate $\Pi$ can be interpreted as the amount of entropy which is being generated 
due to the dissipative dynamics which drives the system towards equilibrium. The entropy flux $\Phi = T^{-1} \frac{d}{dt}H_t$ is related to 
the energy which is dissipated to the environment. The $LS_1$ inequality can in this setting be interpreted 
as a way of bounding the difference between the Free-energies by the entropy production rate. We have that 
\be
	\alpha_1 = \inf_{\rho} \partial_t \log\left(F(\rho_t) -  F_\beta\right),
\ee
where we have that $F(\rho_t) = H_t - TS_t$ as defined above. In other words, for  thermal maps, the $LS_1$ 
constant can be interpreted as the minimal normalized rate of change of the free energy in the system.

\section{Applications}\label{sec:Applications}
In this section we show a number of applications of the abstract results presented in the previous sections. We will only consider examples where $\bL_p$ regularity has been shown to hold (i.e. unital and thermal semigroups).

\subsection{ The depolarizing channel}
To start with, we give an exact expression for the Log-Sobolev constant $\alpha_2$ of the depolarizing semigroup. This is one of the very few cases where it is possible to get an exact explicit expression. 

The generator of the depolarizing semigroup is given by:  
\be \cL(f)=\gamma(\1/d\tr{f}-f)\ee for any $f\in \cA_d$. The corresponding semigroup is easily seen to be 
\be T_t(f)=(1-\epsilon)\1/d\tr{f}+\epsilon f, \ee where $\epsilon=e^{-t\gamma}$.  

\begin{theorem}\label{ThmalphaDepol}
Let $\cL:\cM_d\rightarrow\cM_d$ be the generator of the completely depolarizing semigroup: i.e. $\cL(f)=\gamma(\tr{f}\1/d-f)$ for all $f\in\cA_d$. Then its $LS_2$ constant is given by \be \alpha_2=\frac{2\gamma(1-2/d)}{\log{(d-1)}} \ee
\end{theorem}

\proof{
We show that the result can be obtained by reduction to a lemma proved in \cite{Diaconis}. Recall that 
\be
 \alpha_2=\inf\left\{\frac{\cE_2(f)}{\Ent_2(f)} \; | \; \Ent_2(f)\neq0,f\in\cA_d^+\right\}
 \ee
 Now, noting that the stationary state is $\1/d$, we can write the numerator and denominator explicitly as: 
\be \alpha_2\leq \frac{\tr{f(f-\tr{f}\1/d)}\gamma}{\tr{f(f\log(f)-f\log(||f||_2))}} \label{alphaDepol}\ee

Now, consider an $f\in\cA_d^+$ which saturates Eqn. (\ref{alphaDepol}), then given its spectral decomposition $f=\sum_j f_j\ket{j}\bra{j}$, we get that 

\be f_j\log(f_j)-f_j\log(||f||_2)-\frac{\gamma}{\alpha_2}(f_j-\tr{f}/d)=0, \label{LSfiEqn}\ee for all $j$, as both terms in the sum must be positive. From this point on, the proof mirrors the proof of Theorem A1 in \cite{Diaconis}. In particular, it was shown in  \cite{Diaconis} that the $\{f\}_i$ which saturates Eq. (\ref{LSfiEqn}) is not the uniform distribution.

Now given that $t\rightarrow t\log{t}$ is a convex function, Eqn. (\ref{LSfiEqn}) can only be satisfied for at most two distinct values of $f_j$; call them $x$ and $y$. We know that $f$ cannot be proportional to $\1$, so $f_j=x$ for at least one $j$  and $f_j=y$ also for at least one $j$. Write $d\theta$ for the number of $f_j$ which are equal to $x$, and note that $\theta\in[1/d,1/2]$. Then, by plugging back into Eqn. (\ref{alphaDepol}), we get
\be \alpha_2=\min_{\theta,x,y} \frac{{2}\gamma\theta(1-\theta)(x-y)^2}{\theta x^2\log(x^2)+(1-\theta)y^2\log(y^2)-(\theta x^2+(1-\theta)y^2)\log(\theta x^2+(1-\theta)y^2)}.\ee

This equation can be seen to be exactly the same as the one resulting from the Log-Sobolev constant of a classical projective semigroup with stationary state $\pi=\theta \ket{0}\bra{0}+(1-\theta)\ket{1}\bra{1}$ on a commutative $\bL_p$ space. That problem was solved in \cite{Diaconis}, where it was shown that 
\be \alpha_2=\min_{\theta\in[1/d,1/2]} \frac{{2}\gamma(1-2\theta)}{\log{(1-\theta)\theta}}\ee 
The above minimum is easily seen to be reached for $\theta=1/d$, thus completing the proof.\qed} 

Observe that by l'H\^{o}pitale's rule, we get {$\lim_{d\rightarrow 2}\frac{2(1-2/d)\gamma}{\log(d-1)}=\gamma$.  }

The next natural question which might arise is whether one can evaluate the Log-Sobolev constant of a tensor product of depolarizing semigroups. The Log-Sobolev constant of a classical tensor product semigroup is the minimum of the LS constant of the individual semigroups. In other words, for two classical semigroups $P_t$ and $Q_t$, with Log-Sobolev constants 
$\alpha_2(P_t)$ and $\alpha_2(Q_t)$, we get \cite{Diaconis}
\be \alpha_2(P_t\otimes Q_t)=\min\{\alpha_2(P_t),\alpha_2(Q_t)\} \ee
This however is not guaranteed to be true in the quantum setting because of the possibility for entangled inputs. However, it turns out that it is possible to show this for  qubit depolarizing maps, as illustrated in the following lemma, which was first proved in \cite{Tobias}. We reproduce their proof here in the context of Log-Sobolev inequalities, as it illustrates the power of the hypercontractive method.  

\begin{lemma}\label{qubitDepol}
Let $\gamma>0$, and consider the qubit depolarizing Liouvillian $\cL(f)=\gamma(\tr{f}\1/2- f)$, with $f\in\cA^+_2$. Define the tensor summed Liouvillian on $N$ qubits as 
\be \cL^{(N)}=\cL \otimes\id\otimes ... \otimes \id+\id\otimes\cL \otimes\id\otimes ... \otimes \id+..+\id \otimes ...\otimes\id\otimes \cL\ee
Then,
\be\alpha_2(\cL^{(N)})=\alpha_2(\cL)=\alpha_1(\cL){=\lambda(\cL)=\gamma}\ee\label{2pointineq}
\end{lemma} 

\proof{
This proof relies on certain relationships which were proved for Shatten $p$-norms, therefore we will specify the Shatten $p$- norms by omitting the  $\sigma$  subscript. For unital semigroups, the Shatten $p$ norm and the $\bL_p$ norm are simply related as $||f||_{\sigma,p}=d^{-1/p}||f||_p$, where $f\in\cA_d$ and $\sigma=\1/d$.

 The proof is obtained by induction in the number of tensor powers, and working in the hypercontractive picture. For simplicity of notation we will write $T_t^{(N)}\equiv e^{t\cL^{(N)}}$.

We start by showing that the base case ($N=1$) holds. By theorems \ref{ThmalphaDepol}, and theorem \ref{HypervsLS}, we know that the hypercontractive inequality holds with $\alpha_2\equiv\alpha_2(\cL){=\lambda(\cL)=\gamma}$ for a single tensor power; i.e. $||T^{(1)}_t(f)||_{p,\sigma}\leq||f||_{2,\sigma}$ for all $f\in\cA_2$ and $p\leq1+e^{2t\alpha_2}$. In terms of the Shatten norms this yields $||T_t(f)||_{p}\leq2^{-1/2+1/p}||f||_{2}$. Now assume that $||T^{(N-1)}_t(f)||_{p,\sigma}\leq||f||_{2,\sigma}$ for all $f\in\cA_{2^{N-1}}$ and $p\leq1+e^{2t\alpha_2}$. We will show that the same holds true for $N$ qubits as well.

For some $a,b,c,d\in\cA_{2^{N-1}}$, define $f\in\cA_{2^N}$ and $g\in\cA_2$ as 

\be f=\left( \begin{array}{cc}
a + d & b-ic \\
b+ic & a-d \end{array} \right) ~~~~~~~~~~~ \rm{and} ~~~~~~~~~~~g=\left( \begin{array}{cc}
||a + \epsilon d||_{2,\sigma} & \epsilon||b-ic||_{2,\sigma} \\
\epsilon||b+ic||_{2,\sigma} & ||a-\epsilon d||_{2,\sigma} \end{array} \right),\ee where $\epsilon=e^{-\gamma t}$.

A simple calculation shows that
\be T_t^{(N)}(f)=\left( \begin{array}{cc}
T_t^{(N-1)}(a + \epsilon d) & \epsilon T_t^{(N-1)}(b-ic) \\
\epsilon T_t^{(N-1)}(b+ic) & T_t^{(N-1)}(a - \epsilon d) \end{array} \right)\ee 

Then, for $p\leq \epsilon^{-2}+1$, 
\bq
||T^{(N)}(f)||_{p,\sigma}^p &=& 2^{-N}\lVert \left( \begin{array}{cc}
T_t^{(N-1)}(a + \epsilon d) & \epsilon T_t^{(N-1)}(b-ic) \\
\epsilon T_t^{(N-1)}(b+ic) & T_t^{(N-1)}(a - \epsilon d) \end{array} \right)\rVert_p^p\\
&\leq& \half \lVert \left(\begin{array}{cc}
||T_t^{(N-1)}(a + \epsilon d)||_{2,\sigma} & \epsilon ||T_t^{(N-1)}(b-ic)||_{2,\sigma} \\
\epsilon ||T_t^{(N-1)}(b+ic)||_{2,\sigma} & ||T_t^{(N-1)}(a - \epsilon d)||_{2,\sigma} \end{array}\right)\rVert_p^p\\
&\leq& \half|| \left(\begin{array}{cc}
||a + \epsilon d||_{2,\sigma} & \epsilon ||b-ic||_{2,\sigma} \\
\epsilon ||b+ic||_{2,\sigma} & ||a - \epsilon d||_{2,\sigma} \end{array}\right)||^p_p=||g||_{p,\sigma}^p\eq
where the first inequality follows from a result shown in \cite{King}, and the second inequality follows from hypercontractivity of $T^{(N-1)}_t$ (inductive hypothesis). To complete the proof, we want to show that $||g||_{2,\sigma}\leq||f||_{2,\sigma}$. For that, we will again use hypercontractivity. We define an $h\in\cA_2$ such that $g=T^{(1)}_t(h)$. Then, by hypercontractivity, $||g||_{p,\sigma}\leq||h||_{2,\sigma}$, and finally we will show that $||h||_{2,\sigma}\leq||f||_{2,\sigma}$. 

It is not difficult to see that setting
\bq h_{11}&=&\half((1+\epsilon^{-1})||a+\epsilon d||_{2,\sigma}+(1-\epsilon^{-1})||a-\epsilon d||_{2,\sigma})\\
h_{12}&=&h_{21}=||b-ic||_{2,\sigma}\\
h_{22}&=& \half((1-\epsilon^{-1})||a+\epsilon d||_{2,\sigma}+(1+\epsilon^{-1})||a-\epsilon d||_{2,\sigma})\eq we get that $g=T^{(1)}_t(h)$.

An explicit expansion for $||h||^2_{2,\sigma}$ gives
\be ||h||_{2,\sigma}^2=2||b-ic||_{2,\sigma}^2+(1+\epsilon^{-2})(||a||_{2,\sigma}^2+\epsilon^2||d||_{2,\sigma}^2)+(1-\epsilon^{-2})||a+\epsilon d||_{2,\sigma}||a-\epsilon d||_{2,\sigma},\ee while an explicit expansion for $||f||^2_{2,\sigma}$ yields
\be ||f||_{2,\sigma}^2=(||a||_{2,\sigma}^2+||b||_{2,\sigma}^2+||c||_{2,\sigma}^2+||d||_{2,\sigma}^2)\ee

Hence, in order to complete the proof, we only need to show
\be (1+\epsilon^{-2})(||a||_{2,\sigma}^2+\epsilon^2||d||_{2,\sigma}^2)+(1-\epsilon^{-2})||a+\epsilon d||_{2,\sigma}||a-\epsilon d||_{2,\sigma}\leq 2(||a||_{2,\sigma}^2+||d||_{2,\sigma}^2)\ee
Noting that $(1-\epsilon^{-2})/2$ is negative, it suffices to show
\be (||a||_{2,\sigma}^2-\epsilon^2||d||_{2,\sigma}^2)\leq ||a+\epsilon d||_{2,\sigma}||a-\epsilon d||_{2,\sigma}\ee which follows by the matrix Cauchy-Schwarz inequality.\qed}

\bigskip

To wrap up the discussion on the depolarizing semigroup, we pose a question which was raised in \cite{Pastawski}: is it possible to increase the survival time of a codeword encoded in a system of $N$ qubits suffering local depolarizing noise (with rate $\gamma$), by allowing for arbitrary, possibly time-dependent, Hamiltonian control? In \cite{Pastawski} it was shown that a survival time of order $\log{N}$ can be reached in this manner by conveniently condensing the entropy into specified regions in phase space. Here we show that the upper bound is a direct consequence of the fact that the Dirichlet form for a unital Liouvillian in invariant under the addition of Hamiltonian generators. Indeed, let $\cL'(f)=\cL(f)+i[H,f]$ for some hamiltonian $H'$, then $\cE'_1(f)=\cE_1(f)$ for any $f\in\cA_d$. This furthermore shows that the same argument holds for any unital semigroup, hence extending the statement in \cite{Pastawski} to any primitive unital semigroup. This confirms the intuition that entropy can only be clustered into regions, but can not be eliminated for unital semigroups.

\subsection{Quantum Expanders}

As a second example, we consider the convergence behavior of quantum $D$-regular and expander graphs. There is a vast body of literature on classical expander graphs in the theoretical computer science and combinatorics literature, as these families of graphs have a plethora of useful applications; see \cite{expanderoverview} for a good review. The quantum analogue of this family of graphs has been introduced by several authors, where explicit and implicit constructions have been suggested \cite{qExp1,qExp2,qExp3}. 

A classical $D$-regular graph is a graph where each vertex is connected to exactly $D$ other vertices. A quantum $D$-regular channel is a quantum channel which can be written with exactly $D$ linearly independent Kraus operators. A family of expender graphs (where the dimension specifies the elements in the family) is a set of $D$-regular graphs such that the spectral gap is asymptotically independent of the dimension. A family of quantum expander channels is analogously a set of  $D$-regular channels such that the spectral gap of the channels is asymptotically independent of the dimension.
Expanders graphs are often used as efficient randomness generators, where one considers a random walk on the expander, and because of the constant spectral gap, the initial population spreads evenly across the graph very rapidly. 

The main theorem of this section shows that the Log-Sobolev constant can be qualitatively different from the spectral gap, and provides a much more informative upper bound on the convergence of quantum expanders. It is important to point out that quantum expanders are defined as (time-) discrete channels, whereas the Log-Sobolev tools were developed for continuous time semigroups. However, given a primitive quantum channel $T:\cM_d\rightarrow \cM_d$ we can define the Liouvillian $\cL= T-\id$ and relate their spectra and mixing properties. This correspondence is outlined in lemma \ref{Lem:DCtime}.
We therefore define the Log-Sobolev constant of the channel $T$ as the Log-Sobolev constant of the associated  Liovillian  $\cL= T-\id$. 

\begin{theorem}\label{Thm:expanders}
The Log-Sobolev constant of $\cL= T-\id$, where $T:\cM_d\rightarrow\cM_d$ is any $D$-regular reversible unital channel  satisfies:
\be \alpha_2 \leq \log{D}\frac{4+\log{\log{d}}}{2\log{3d/4}}\ee
\end{theorem}

\proof{
Given the reversible $D$-regular unital channel $T$, consider the lazy channel defined as $\tilde{T}=\half(\id+T)$. Associate to this lazy channel a lazy Liouvillian $\tilde{\cL}\equiv \tilde{T}-id$. For an initial pure state $\varphi$, it is clear that the rank of the output state of a $D$-regular channel will be at most $D^n$ after $n$ iterations of the map. Let $P_n$ be the projector onto the complement of the support of $\tilde{T}^n(\varphi)$ (i.e. $P_n\tilde{T}^n(\varphi)P_n=0$). Then, given that $\tr{X}\geq\tr{P_nXP_n}$ for any $X\in\cA_d^+$, 

\bq \chi^2(\tilde{T}^n(\varphi),\1/d)&=&\tr{(\tilde{T}\circ\Gamma^{-1/2}(\varphi)-\1/\sqrt{d})^2}\\
&\geq&\tr{P(\tilde{T}\circ\Gamma^{-1/2}(\varphi)-\1/\sqrt{d})^2P}\\
&=& \tr{P}/d\geq\frac{d-D^n}{d}\eq

This implies that $\chi^2(\tilde{T}^n(\varphi),\1/d)\geq 1/2$ whenever $D^n/d\leq 3/4$; or $n\leq \frac{\log(3d/4)}{\log(D)}$. An upper bound on the $\chi^2$ divergence can be obtained by combining lemma \ref{Lem:DCtime} and proposition \ref{chi2gapbound}(proved later on in this section). Indeed, lemma \ref{Lem:DCtime} guarantees that we can upper bound the $\chi^2$ divergence of the channel $\hat{T}$ by the $\chi^2$ divergence of the semigroup with Liouvillian $\tilde{\cL}\equiv \tilde{T}-id$, while proposition \ref{chi2gapbound} guarantees that 
\be \chi^2(T^n(\varphi),\1/d) \leq \half\label{Eqnchi2ub}\ee when $n\geq \frac{1}{2\alpha_2}(\log\log(d)+1)$. In deriving Eqn. (\ref{Eqnchi2ub}), we have also used that $\alpha_2\leq\lambda$, that $\tilde{\alpha}_2=2\alpha_2$ and that $2-\log(2)\leq1$. Finally, combining the upper and the  lower bounds on the $\chi^2$ divergence yields:
\be\alpha_2\leq \log{D}\frac{4+\log{\log{d}}}{2\log{3d/4}}\ee and completes the proof. \qed}

\textbf{Note}: This method cannot be used to bound the $LS_1$ constant, as it crucially depends on the correspondence between Log Sobolev inequalities and hypercontractivity in the proof of proposition \ref{chi2gapbound}. 

\bigskip

theorem \ref{Thm:expanders} holds quite independently of the actual scaling of the spectral gap or of the Log Sobolev constant. It simply gives an absolute upper bound on $\alpha_2$ in terms of the Kraus rank and of the dimension for primitive reversible unital channels. This upper bound however becomes particularly relevant in the context of expanders. It shows that for expanders, even though the spectral gap is asymptotically independent of the dimension $d$, the $LS_2$ constant will always decrease logarithmically with the dimension. We corroborate this claim by providing a general lower bound on $\alpha_2$ in terms of the spectral gap for unital semigroups. 

\begin{corollary}
Let $\cL:\cM_d\rightarrow\cM_d$ be a primitive unital Liouvillian with spectral gap $\lambda$, then  
\be \alpha_2\geq\frac{{2}(1-2/d)\lambda}{\log{(d-1)}}\ee
\end{corollary}

\proof{
Note that for $\cL^{\rm depol}(f)\equiv(\tr{f}\1/d-f)$, then for any $f\in\cA$ we get that $\Var_\sigma(f)=\cE^{\rm depol}_2(f)$. Hence, if $\cE_2(f)$ is the Dirichlet form associated with $\cL$, then
\be \frac{{2}(1-2/d)}{\log{(d-1)}}\Ent_2(f)\leq \cE^{\rm depol}_2(f)=\Var_\sigma(f)\leq \cE_2(f)/\lambda\ee where the first inequality is obtained from theorem \ref{ThmalphaDepol}, and the last one follows from the variational characterization of the spectral gap  in Eqn. (\ref{spectGapVar}).\qed}

Thus, combining the two bounds, we get that for any primitive, reversible and unital $D$-regular Liouvillian $\cL:\cM_d\rightarrow\cM_d$,  with Log-Sobolev constant $\alpha_2$ and spectral gap $\lambda$:

\be \frac{{2}(1-2/d)\lambda}{\log{(d-1)}}\leq \alpha_2\leq \log{D}\frac{4+\log{\log{d}}}{2\log{3d/4}} \ee

In particular, lemma \ref{Lem:DCtime}, and the mixing time analysis in section \ref{sec:Mixing}, provide further evidence that the mixing time of a quantum expander cannot in general terms be faster than $\cO(\log(d))$.
It is worth mentioning that a very important class of expander channels, namely random unitary channels \cite{qExp3} are unital and reversible.

\bigskip

The remainder of this section consists of the lemmas which were used in the proof of theorem \ref{Thm:expanders}. We will first need a lemma relating the $\chi^2$ mixing of a channel $T:\cM_d\rightarrow\cM_d$ to that of its associated semigroup $\cL\equiv(T-id)$:

\begin{lemma}\label{Lem:DCtime}
Let $T:\cM_d\rightarrow\cM_d$ be a primitive reversible quantum channel with stationary state $\sigma$, and let $T_t$ be the semigroup with generator $\cL=T-\id$. Moreover, suppose that $T$ (and consequently $T_t$) is lazy; i.e. there exists a quantum channel $S$ such that $T=\half(id+S)$. Then, for any positive integer $n$, and any input state $\rho\in\cS_d$,
\be \chi^2(T^{*n}(\rho),\sigma)\leq \chi^2(T^*_n(\rho),\sigma),\ee where $T_n$ refers to the continous time semigroup, and $T^n$ refers to discrete powers of the quantum channel.
\end{lemma}
\proof{
Given that $T$ is reversible, its spectrum is real and we can define the similarity transform $\tilde{T}\equiv \Gamma_\sigma^{1/2}\circ T\circ\Gamma_\sigma^{-1/2}$. This map is Hermitian and  possesses an orthonormal basis of eigen-operators; write them as $\{E_k\}$. Note that both maps have the same spectrum, as they are related by a similarity transformation. Primitivity guarantees that the second largest eigenvalue is strictly smaller than $1$, and laziness ensures that the spectrum is non-negative. We therefore can write the eigenvalues $\{\beta_i\}$ of $\tilde{T}$ in decreasing order as $1=\beta_0>\beta_1\geq ...\geq \beta_{d^2-1}\geq0$. We similarly write the eigenvalues $\{\lambda_i\}$ of $\tilde{\cL}=\gamma(\tilde{T}-\id)$ as $\lambda_i\equiv\beta_i-1$ for $i=1,...,d^2-1$. 
Under these assumptions, $\tilde{T}$ can be written as
\be \tilde{T}(\rho)=\sum_{k=0}^{d^2-1}\beta_k \tr{E_k^\dag\rho}E_k\ee where $E_0=\sqrt{\sigma}$. Then
\bq \chi^2(T^{*n}(\rho),\sigma)&=&\tr{(T^{*n}(\rho)-\sigma)\Gamma_\sigma^{-1}(T^{*n}(\rho)-\sigma)}\\
&=& \tr{|\tilde{T}^n\circ\Gamma_\sigma^{-1/2}(\rho)-\sqrt{\sigma}|^2}\\
&=& \tr{|\sum_{k=1}^{d^2-1}\beta^n_{k}\tr{E^\dag_k\Gamma^{-1/2}_\sigma(\rho)}E_k|^2}\label{Eqn142}\eq

Now, note that the coefficients in Eqn. (\ref{Eqn142}) are all positive. Then, since $\log{(1+x)}\leq x$ for all $x\in[0,1]$, it follows that 
\be \beta_i=e^{\log{(1+\lambda_i)}}\leq e^{\lambda_i}\ee

Thus,
\be \chi^2(T^{*n}(\rho),\sigma) \leq \tr{|\sum_{k=1}^{d^2-1}e^{n\lambda_{k}}\tr{E^\dag_k\Gamma^{-1/2}_\sigma(\rho)}E_k|^2} = \chi^2(T^*_n(\rho),\sigma)\ee\qed}

In order to simplify the presentation of what follows, we introduce the notation of $p\rightarrow q$ norms. We will write $||T_t||_{(q,\sigma)\rightarrow(p,\sigma)}\leq 1$ to mean that $||T_t(f)||_{p,\sigma}\leq||f||_{q,\sigma}$ for all $f\in \cA_d$. 
We additionally introduce two lemmas, in order to clarify the proof of the next theorem:  

\begin{lemma}\label{dualityLp}
Let $T:\cM_d\rightarrow\cM_d$ be a primitive quantum channel with stationary state $\sigma\in\cS_d^+$, then 
\be||T||_{(p,\sigma) \rightarrow (q,\sigma)}\leq 1 ~~~~~ \rm{iff.} ~~~~~||\hat{T}||_{(q',\sigma) \rightarrow (p',\sigma)}\leq 1,\ee where $p',q'$ are the H\"older duals of $p,q$, and $\hat{T}=\Gamma_\sigma^{-1}\circ T^*\circ\Gamma_\sigma$.
\end{lemma}
\proof{ 
The proof is a straightforward consequence of the duality of $\bL_p$ norms (lemma 1.3). Assume that $||T||_{(p,\sigma) \rightarrow (q,\sigma)}\leq 1$, then for any $f\in\cA_d$, $||T(f)||_{q,\sigma}\leq||f||_{p,\sigma}$. Now let $g\in\cA_d$, and consider
\bq ||\hat{T}(g)||_{p',\sigma} &=& \sup \{\avr{f,\hat{T}(g)}_\sigma, ||f||_{p,\sigma},f\in\cA_d\}\\
 &=& \sup\{\avr{T(f),g}_\sigma, ||f||_{p,\sigma},f\in\cA_d\}\\
 &=& \sup\{\avr{T(f),g}_\sigma, ||T(f)||_{p,\sigma},f\in\cA_d\}\\
 &\leq& \sup\{\avr{h,g}_\sigma, ||h||_q,h\in\cA_d\}\\ &=& ||f||_{q',\sigma}\eq
The proof of the other direction proceeds in exactly the same manner.\qed}

\textbf{Note}: by a slight modification of the argument in \cite{Watrous}, it can be shown that the supremum in $||T(f)||_{q,\sigma}\leq||f||_{p,\sigma}$ is reached for some positive matrix $f\in\cA_d^+$.

\begin{lemma}\label{lemma2to2norm}
Let $\cL:\cM_d\rightarrow\cM_d$ be a primitive Liouvillian with spectral gap $\lambda$, and let $T_t$ be its associated semigroup. Then, 
\be ||T_t-T_\infty||_{(2,\sigma)\rightarrow (2,\sigma)}\leq e^{-t\lambda},\ee where $T_\infty=\lim_{t\rightarrow\infty}T_t$.
\end{lemma}
\proof{
We start by recalling the variational characterization of the spectral gap:
\bq \lambda&=&\min_{f\in\cA_d}\left \{\frac{\cE_2(f)}{\Var_\sigma(f)},\Var(f)\neq0 \right \}\\
&=&\min_{f\in\cA_d} \left \{\cE_2(f), ||f||_{2,\sigma}=1,\tr{\Gamma_\sigma(f)}=0 \right \}\eq where we have used that $\Var_\sigma(f)=||f||_{2,\sigma}^2-\tr{\Gamma_\sigma(f)}$. Now, observe that for any $f\in\cA_d$, $\Var_\sigma(f)$ and $\cE_2(f)$ are invariant under the transformation $f\mapsto f+c\1$ with $c\in\bR$. Therefore, we can without loss of generality assume that $f\in\cA_d^+$. Then,  
\be \partial_t||(T_t-T_\infty)(f)||_{2,\sigma}^2=-2\cE_2(T_t(f))\leq-2\lambda \Var_\sigma(T_t(f))=-2\lambda||(T_t-T_\infty)(f)||_{2,\sigma}^2\ee
Thus, 
\be ||(T_t-T_\infty)(f)||_{2,\sigma}^2\leq e^{-2\lambda t}\Var_\sigma(f)\ee
Taking the supremum over $f\in\cA_d$ such that $||f||_2=1$ and $\tr{\Gamma_\sigma(f)}=0$, then completes the proof.\qed}

We are now in a position to prove the main technical result of this section which is an inequality involving the $\chi^2$-divergence for primitive Liouvillians.

\begin{proposition}\label{chi2gapbound}
Let $\cL:\cM_d\rightarrow\cM_d$ be a primitive Liouvillian with Log-Sobolev  constant $\alpha_2$ and spectral gap $\lambda$, and let $T_t$ be its associated semigroup. Let $\sigma_{\min}$ denote the smallest eigenvalue of $\sigma$. Then, for any $c>0$ and $t\geq\frac{1}{2\alpha_2}(\log{\log{1/\sigma_{\min}}})+\frac{c}{\lambda}$, 
\be \chi^2(T^*_t(\rho),\sigma)\leq e^{2(1-c)}\ee for any state $\rho\in\cS_d$.
\end{proposition}
\proof{
Let $p(t)=1+e^{2\alpha_2 t}$. By theorem \ref{HypervsLS} we get hypercontractivity of the semigroup: $||T_t||_{(2,\sigma)\rightarrow (p(t),\sigma)}\leq 1$. Then by lemma \ref{dualityLp}, we also get that $||\hat{T}_t||_{(p'(t),\sigma)\rightarrow(2,\sigma)}\leq1$, where $1/p+1/p'=1$, and $\hat{T_t}=\Gamma^{-1}_\sigma T^*_t \Gamma_\sigma$. Now, writing $\rho_{t+s}\equiv T^*_{t+s}(\rho)$, for any $\rho\in\cS_d$ and positive reals $s,t>0$, we get

\bq \sqrt{\chi^2(\rho_{t+s},\sigma)} &=& \sqrt{\tr{(\rho_{t+s}-\sigma)\Gamma^{-1}_\sigma(\rho_{t+s}-\sigma)}} \\
&=& ||\hat{T}_{t+s}(\Gamma^{-1}_\sigma(\rho))-\sigma||_{2,\sigma}\\
&=& ||(\hat{T}_{t+s}-\hat{T}_\infty)(\Gamma^{-1}_\sigma(\rho))||_{2,\sigma}\\
&\leq& ||\hat{T}_s(\Gamma_\sigma^{-1}(\rho))||_{2,\sigma}||\hat{T}_t-\hat{T}_\infty||_{(2,\sigma)\rightarrow(2,\sigma)}\\
&\leq& ||\hat{T}_s||_{(p'(s),\sigma)\rightarrow (2,\sigma)}||\Gamma^{-1}_\sigma(\rho)||_{p'(s),\sigma}||\hat{T}_t-\hat{T}_\infty||_{(2,\sigma)\rightarrow(2,\sigma)}\\
&\leq& \sigma_{\min}^{-1/p(s)}e^{-t\lambda}||\hat{T}_s||_{(p'(s),\sigma)\rightarrow (2,\sigma)}\\
&\leq& \sigma_{\min}^{-1/p(s)}e^{-t\lambda},\eq where the inequalities follow from hypercontractivity of $T_t$, lemma \ref{lemma2to2norm}, and properties of the $\bL_q$ norms (lemma \ref{Lem:Lp-norm}). Note in particular, that for a given $t>0$, $\hat{T}_t$ is a quantum channel, and that it has the same spectrum as $T_t$.  

Choosing $s=\frac{1}{2\alpha_2}(\log{\log{1/\sigma_{\min}}})$, we get $p(s)=1-\log{\sigma_{\min}}$. By noting that $1/p\geq1/(1+p)$, we get
\be \chi^2(T^*_t(\rho),\sigma)\leq e^{1-\lambda t}, \ee which completes the theorem.\qed}

\section{Outlook}\label{sec:Conclusion}
In this paper, we have introduced the tools of logarithmic Sobolev inequalities for the analysis of mixing times of quantum dynamical semigroups. We have identified the relevant Log-Sobolev constants in the case of finite state spaces, and proved upper bounds in terms of the spectral gap of the generator of the semigroup. We have recast the well-known equivalence between Log-Sobolev inequalities and Hypercontractivity in the finite non-commutative state space setting, and have shown that the equivalence carries over essentially unchanged from the classical case if the Liouvillian satisfies an $\bL_p$ regularity condition. We show that unital and thermal (Davies generators) Liouvillians satisfy such a condition.

Having worked out the abstract theory, we showed that it implies very strong bounds on the mixing time of the semigroup, when the spectral gap and the Log-Sobolev constant are comparable. In particular, the pre-factor associated with the mixing is exponentially smaller than the one obtained by a $\chi^2$ bound. We have explicitly calculated the Log-Sobolev constant for the depolarizing channel of dimension $d$, and of a tensor product of qubit depolarizing channels. Finally, we have provided upper and lower bounds on the Log-Sobolev constant of $D$-regular unital channels, and discussed implications for quantum expanders. In particular, we showed that even though the gap of a random unitary channel is asymptotically independent of dimension, its Log-Sobolev constant will decrease as $\cO(1/\log(d))$. 

To conclude, we briefly discuss potential further application of the framework introduced here, as well as issues which have been left unresolved.

\begin{itemize}
\item We have to point out,  that we have to a large extent  only introduced the formal setting of Log-Sobolev inequalities, and that many relevant applications remain to be worked out. In the classical setting, Log-Sobolev inequalities and hypercontractivity have been extremely useful tools. One area where they have proved to be paramount is in analyzing the mixing properties of spin systems on a lattice under Glauber dynamics. Several authors have been able to show a number of very tight mixing results \cite{Martinelli,Martinelli2}, in particular relating spacial and temporal mixing in a one-to-one fashion. It would be very desirable to generalize these results to the quantum setting. More generally, a number of methods, including block renormalization transformations and comparison theorems, have been developed in the classical setting in order to explicitly calculate the $LS_2$ constant for specific systems. It would be very important to generalize these results to the quantum setting. 
\item There also remain a number of open questions in the abstract theory of quantum Log-Sobolev inequalities. We mention two which we consider important to resolve. The first is to settle Conjecture \ref{Lpconjecture}. On the one hand, it would be interesting to know whether there exist semigroups which violate $\bL_p$ regularity, as it would be a distinctly quantum signature in the theory. Conversely, if the conjecture is true then the Log-Sobolev machinery can be used quite generally, and one can expect that most of the classical mixing time tools can be inherited with little modification. The second open question is to figure out whether the Log Sobolev constants of a tensor power of semigroups is equal to the Log Sobolev constant of its components. This problem can be rephrased in several different ways which could have relevance in quantum information theory.  It can, for instance, be related in a one-to-one manner to various forms of multiplicativity of $2\rightarrow p$ norms, and as such, provides an important operational interpretation for these quantities. As far as we know, this is an open question in operator space theory. 

\end{itemize}

\textbf{Acknowledgements} We thank Frank Verstraete and Michael Wolf for insightful discussions. 
MJK acknowledges support from the Alexander-von-Humboldt Foundation and from the Niels Bohr International Academy. KT is grateful for the support from the Erwin Schr\"odinger fellowship, Austrian Science Fund (FWF): J 3219-N16. This work was also supported by the EU projects (QUEVADIS,Q-Essence), and the BMBF (QuOReP).

\bibliographystyle{unsrt}

\end{document}